\documentclass[10pt,final,doublecolumn]{IEEEtran}
\usepackage{subfigure}
\usepackage{amsmath}
\usepackage{latexsym}
\usepackage{graphicx}
\usepackage{bbding}
\usepackage{indentfirst}
\usepackage{setspace}
\usepackage{cite,balance}
\usepackage{float}
\usepackage{epstopdf}
\usepackage{amssymb}
\usepackage{amsfonts}
\usepackage{enumerate}
\usepackage{multicol}
\usepackage{color}
\usepackage{stfloats}
\usepackage{bm}
\usepackage{amssymb}
\usepackage{stfloats}
\usepackage{algorithm,algpseudocode,float}
\usepackage{lipsum}
\usepackage{makecell}
\usepackage{array}
\usepackage[table]{xcolor}
\usepackage{makecell}
\usepackage{pifont}
\IEEEoverridecommandlockouts
\allowdisplaybreaks[4]



\begin{document}
\title{Target-Mounted Intelligent Reflecting Surface for Secure Wireless Sensing}
\author{{Xiaodan Shao,~\IEEEmembership{Member,~IEEE}, and Rui Zhang, \IEEEmembership{Fellow, IEEE}}
	 \thanks{X. Shao is with the School of Science and Engineering, Chinese University of Hong Kong, Shenzhen, China 518172, (e-mail: shaoxiaodan@zju.edu.cn).}

\thanks{R. Zhang is with School of Science and Engineering, Shenzhen Research Institute of Big Data, The Chinese University of Hong Kong, Shenzhen, Guangdong 518172, China. He is also with the Department of Electrical and Computer Engineering, National University of Singapore, Singapore 117583 (e-mail: elezhang@nus.edu.sg).
%\emph{(Corresponding author: Rui Zhang)}
}
}
\maketitle
\begin{abstract}
In this paper, we consider a challenging secure wireless sensing scenario where a legitimate radar station (LRS) intends to detect a target at unknown location in the presence of an unauthorized radar station (URS). We aim to enhance the sensing performance of the LRS and in the meanwhile prevent the detection of the same target by the URS. Under this setup, conventional stealth-based approaches such as wrapping the target with electromagnetic wave absorbing materials are not applicable, since they will disable the target detection by not only the URS, but the LRS as well. To tackle this challenge, we propose in this paper a new target-mounted IRS approach, where intelligent reflecting surface (IRS) is mounted on the outer/echo surface of the target and by tuning the IRS reflection, the strength of its reflected radar signal in any angle of departure (AoD) can be adjusted based on the signal's angle of arrival (AoA), thereby enhancing/suppressing the signal power towards the LRS/URS, respectively. To this end, we propose a practical protocol for the target-mounted IRS to estimate the LRS/URS channel and waveform parameters based on its sensed signals and control the IRS reflection for/against the LRS/URS accordingly. Specifically, we formulate new optimization problems to design the reflecting phase shifts at IRS for maximizing the received signal power at the LRS while keeping that at the URS below a certain level, for both the cases of short-term and long-term IRS operations with different dynamic reflection capabilities. To solve these non-convex problems, we apply the penalty dual decomposition method to obtain high-quality suboptimal solutions for them efficiently. Finally, simulation results are presented that verify the effectiveness of the proposed protocol and algorithms for the target-mounted IRS to achieve secure wireless sensing, as compared with various benchmark schemes.
\end{abstract}


\IEEEpeerreviewmaketitle

\section{Introduction}
The future sixth-generation (6G) wireless networks are envisioned as an enabler for various emerging applications, such as extended reality, tactile internet, intelligent transportation, massive Internet of Things (IoT), etc \cite{6G}. These applications generally require extraordinarily high communication rate and/or reliability, as well as highly accurate sensing. Moreover, with the growing demand for wireless communication system capacity, spectrum/hardware resources are becoming increasingly scarce. The above trends have motivated a new research paradigm, called integrated sensing and communications (ISAC), for enabling sensing and communication functions in the same wireless network by sharing its platform and resources such as base station (BS), antenna, spectrum, waveform, etc. and jointly designing their operations to achieve balanced performance \cite{zeng_wave, liuan,cui,use}. It is thus anticipated that wireless sensing, which aims to accurately and efficiently detect, estimate, and extract useful physical information/features of environmental targets by exploiting radio wave transmission, reflection, diffraction, and scattering, will become a major service provided by 6G wireless networks, in addition to communications.

However, existing wireless sensing systems such as cellular BS sensing still face practical challenges. For example, the sensing range/accuracy of the radar (BS) is practically limited due to the high round-trip radar signal propagation loss between the radar and its sensing target as well as the target's finite radar cross section (RCS). Even under the line-of-sight (LoS) condition, when the radar-target  distance is long and/or the RCS of the target is small, the echo signal received by the radar receiver is of low power, which degrades the sensing performance (e.g., target detection probability) against the background noise. Although this issue can be partially resolved by increasing the
radar signal power and/or employing multiple-input-multiple-output (MIMO) radars, such conventional methods usually require substantially more energy  consumption and higher implementation cost.

In recent years, intelligent reflecting surface (IRS) has been proposed as a promising new technology to achieve smart and reconfigurable signal propagation for wireless communications \cite{qing,wei}. Typically, IRS is a planar surface composed of a large number of low-cost passive reflecting elements, and by jointly adjusting the amplitude and/or phases of IRS elements, the direction and strength of the electromagnetic wave reflected by IRS can be flexibly  controlled (i.e., passive beamforming). Motivated by the significant performance gains that IRS brings for wireless communications, recent studies have also revealed the great potential of IRS for enhancing the performance of wireless sensing \cite{xujie,s2,s3,fangjun} as well as improving the communication-sensing trade-off for ISAC systems \cite{isac0, isac-2,isac-3}. In these works, IRS is employed as additional anchor node in wireless network for improving the BS's detection accuracy, which, however, may not perform well in practice due to the significant path loss of the radar signal that is reflected by both the target (once) and the IRS (twice) before being received by the radar/BS receiver. To overcome this issue, in \cite{shaos} and \cite{shaos1}, the authors proposed a new IRS active sensing approach, where the
IRS (instead of cellular BS) directly sends radar signal, and sensors
installed on the IRS receive the echo reflected by the target
for detection, thus avoiding the severe (double)
path loss between the sensing/ISAC BS
and IRS in conventional IRS-aided cellular sensing systems.

On the other hand, security is another critical challenge for wireless sensing due to the inherent broadcast nature of wireless signals. Since IRS
has the capability of enhancing wireless
signals at desired receivers as well as suppressing them at undesired receivers, it has been applied to improve the communication security in IRS-aided ISAC systems \cite{qings,seuliu}. However, prior works only focused on IRS reflection design to achieve enhanced 
{\it information/communication} security in ISAC systems, but overlooked the important {\it sensing security} threat. In practice, the physical characteristics of sensing target (e.g., location and velocity) may be intercepted by unauthorized radar stations (URSs), which thus brings a new sensing security issue. There are two traditional methods for tackling this issue. One method is by designing the target's shape \cite{shape}, such that it can reflect incident signal to undetectable directions, while this method has limited practical applications due to the specific target shape required. The other method is to wrap the target with  electromagnetic (EM) wave absorbing (i.e., EM stealthy) materials \cite{stealth}, which can significantly reduce the reflected radar signal power in all directions. Although the above methods can achieve sensing security against URS, they inevitably render the target invisible to any legitimate radar station (LRS) as well. In addition, other methods have been recently proposed to enable the LRS to detect stealthy target such as increasing the power-aperture product of the LRS, or the number of LRS radar pulses in coherent processing \cite{PA}. However, these methods can also be applied by the URS to enhance its target detection. Note that the active IRS sensing approach \cite{shaos,shaos1} aforementioned cannot solve the secure sensing problem too, because it has no control over the reflected signal by the target.
\begin{figure}[t]
\setlength{\abovecaptionskip}{-0.cm}
\setlength{\belowcaptionskip}{0.cm}
  \centering
\includegraphics [width=0.49\textwidth] {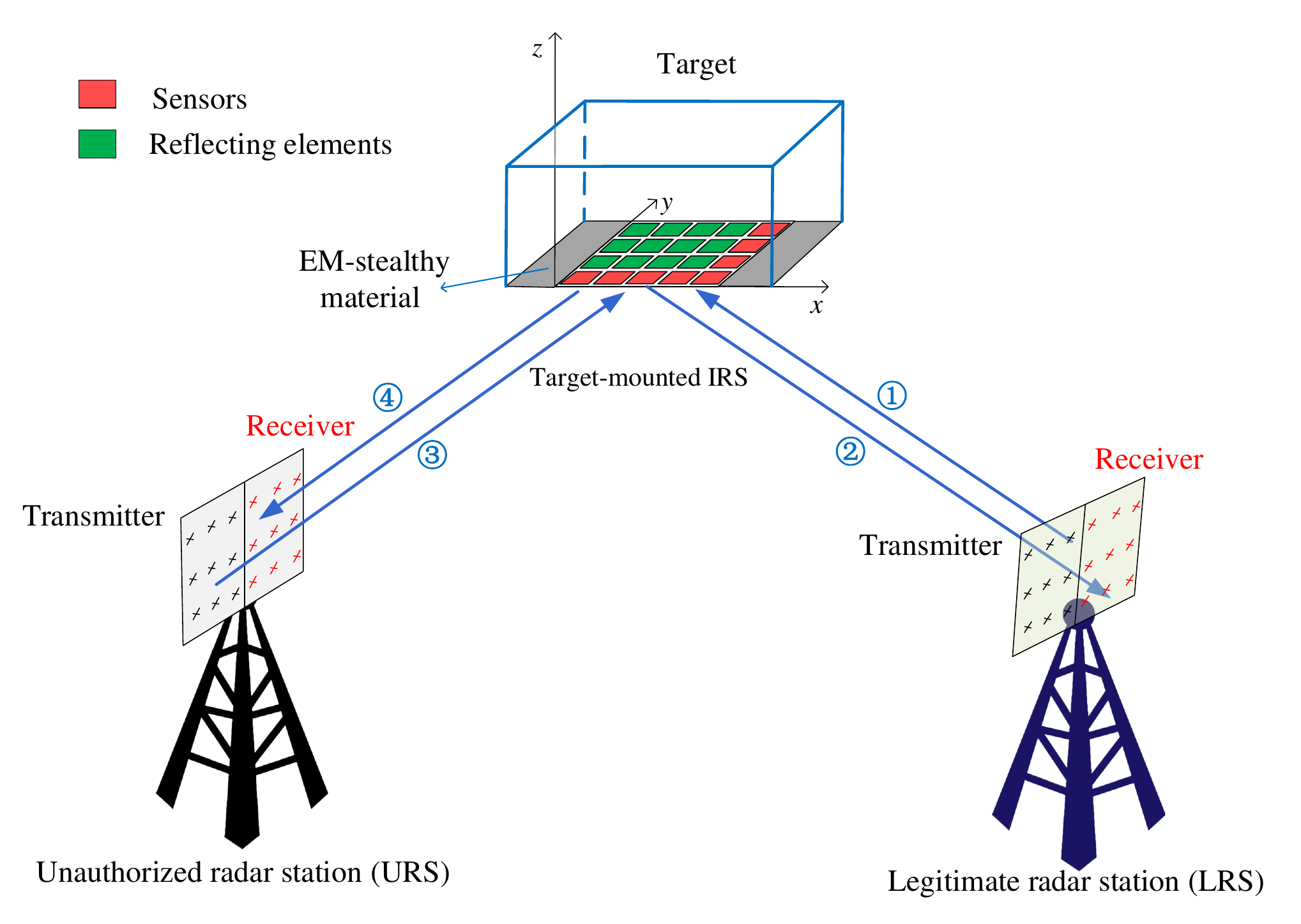}
\caption{System model of secure wireless sensing with target-mounted IRS.}
\label{sys_model}
\end{figure}

In order to improve the target sensing performance while ensuring sensing security at the same time, we propose in this paper a new approach by mounting IRS on the outer surface of the target, namely {\it target-mounted IRS}. Specifically, we consider a secure wireless sensing system as shown in Fig. \ref{sys_model}, where an LRS aims to detect an aerial target with its echo-surface covered by an IRS, and a URS also intends to detect the same target. The target-mounted IRS can help achieve secure sensing of the target for the LRS as well as against the URS. Specifically, instead of using IRS as additional anchor node in existing IRS-assisted sensing/ISAC systems, mounting IRS on the target can directly control the reflected echo signal from the target. By tuning IRS reflection, the echo signal towards the LRS/URS receiver can be greatly enhanced/suppressed, thus achieving our secure wireless sensing goals. Note that the above IRS reflection design generally replies on the knowledge of the angles of arrival (AoAs) of the radar signals from the LRS and/or URS. To achieve this end, we consider installing dedicated sensors along with the reflecting elements at the IRS to enable its acquisition of such information (see Fig. \ref{sys_model}). The main contributions of this paper are summarized as follows.
\begin{itemize}
\item
First, we propose a practical operation protocol for target-mounted IRS to achieve secure wireless sensing. The protocol consists of two steps. In the first step, IRS sensors estimate the LRS/URS channel and waveform parameters with all IRS reflecting elements switched off. Then, in the second step, based on the estimated parameters, IRS reflection is designed to simultaneously boost the received signal at the LRS receiver and suppress that at the URS receiver, thus achieving secure detection of the target.

\item
Next, to design the IRS reflection for/against the LRS/URS, we formulate new optimization problems for both the cases of short-term and long-term IRS operations with different dynamic reflection capabilities, aiming to maximize the received signal power at the LRS while in the meanwhile keeping that at the URS below a given threshold. However, the formulated optimization problems are non-convex and thus difficult to be solved optimally. To overcome this difficulty, we propose an efficient penalty dual decomposition (PDD)-based algorithm to solve these problems, which yields high-quality suboptimal solutions.

\item
Finally, we evaluate the performance of our proposed
designs via numerical results. The results demonstrate that target-mounted IRS can greatly improve the target's detection accuracy by the LRS in the LRS-only scenario, or degrade the detection performance by the URS in the URS-only scenario, as well as simultaneously enhance/degrade the target sensing at LRS/URS when they are both present. In addition, it is shown that the proposed algorithm achieves comparable performance to the semi-definite relaxation (SDR)-based method, but with much lower computational complexity required, thus is more suitable for real-time implementation. Furthermore, it is shown that the target-mounted IRS with proposed reflection optimization achieves more robust secure sensing performance against IRS sensing errors and random LRS/URS locations, as compared to benchmark systems without IRS or with random IRS reflection.
\end{itemize}

The rest of this paper is organized as follows. Section II
presents the channel and signal models of the proposed secure wireless sensing system with target-mounted IRS. Section III describes the  practical protocol for the operation of target-mounted IRS.
Section IV formulates
the optimization problems for IRS reflection design in both the short-term and long-term IRS reflection cases, and proposes efficient algorithms for solving them. Section V presents numerical results to evaluate
the performance of the proposed protocol and algorithms. Finally,
Section VI concludes this paper.

\emph{Notations}: Boldface upper-case and lower-case letters denote
matrices and vectors, respectively, $(\cdot)^*$, $(\cdot)^H$, and $(\cdot)^T$  respectively denote conjugate, conjugate transpose, and transpose, $\otimes$ denotes the Kronecker product, $\odot$ denotes the Hadamard product, $\mathbb{E}[\cdot]$ denotes the expected value of
random variable, $\left \|\cdot\right \|_2$ denotes the Euclidean norm, $|\cdot|$ denotes the absolute value, $\Re(\cdot)$ denotes the real part of a complex number, $\mathbf{I}_{m}$ denotes the $m$-dimensional identity matrix, $\mathrm{diag}({\bf x})$ denotes a diagonal matrix with the diagonal entries specified by a vector ${\bf x}$.

\section{System Model}
In this paper, we consider a secure wireless sensing system aided by the target-mounted IRS as illustrated in Fig. \ref{sys_model}, where an LRS aims to detect a target at unknown location, while a URS also intends to detect  the same target by sending/intercepting radar signals. Specifically, we consider that an IRS is equipped on the echo surface of the target to help boost/suppress its reflected signal towards the LRS/URS, which are both assumed to be a mono-static radar (i.e., with the radar transmitter and receiver co-located). In addition, we assume that the remaining echo surface of the target, which is uncovered by the IRS, is
wrapped with EM-stealth material \cite{stealth}, such that the IRS can completely control the reflected radar signal from the target\footnote{Note that the target is assumed to be passive and thus it cannot emit signals to the LRS for facilitating its detection, due to various practical considerations (e.g., information/sensing security, energy/hardware  constraint).}.

\subsection{Channel Model}
\begin{figure}[t!]
\setlength{\abovecaptionskip}{-0.cm}
\setlength{\belowcaptionskip}{0.cm}
  \centering
\includegraphics [width=0.51\textwidth] {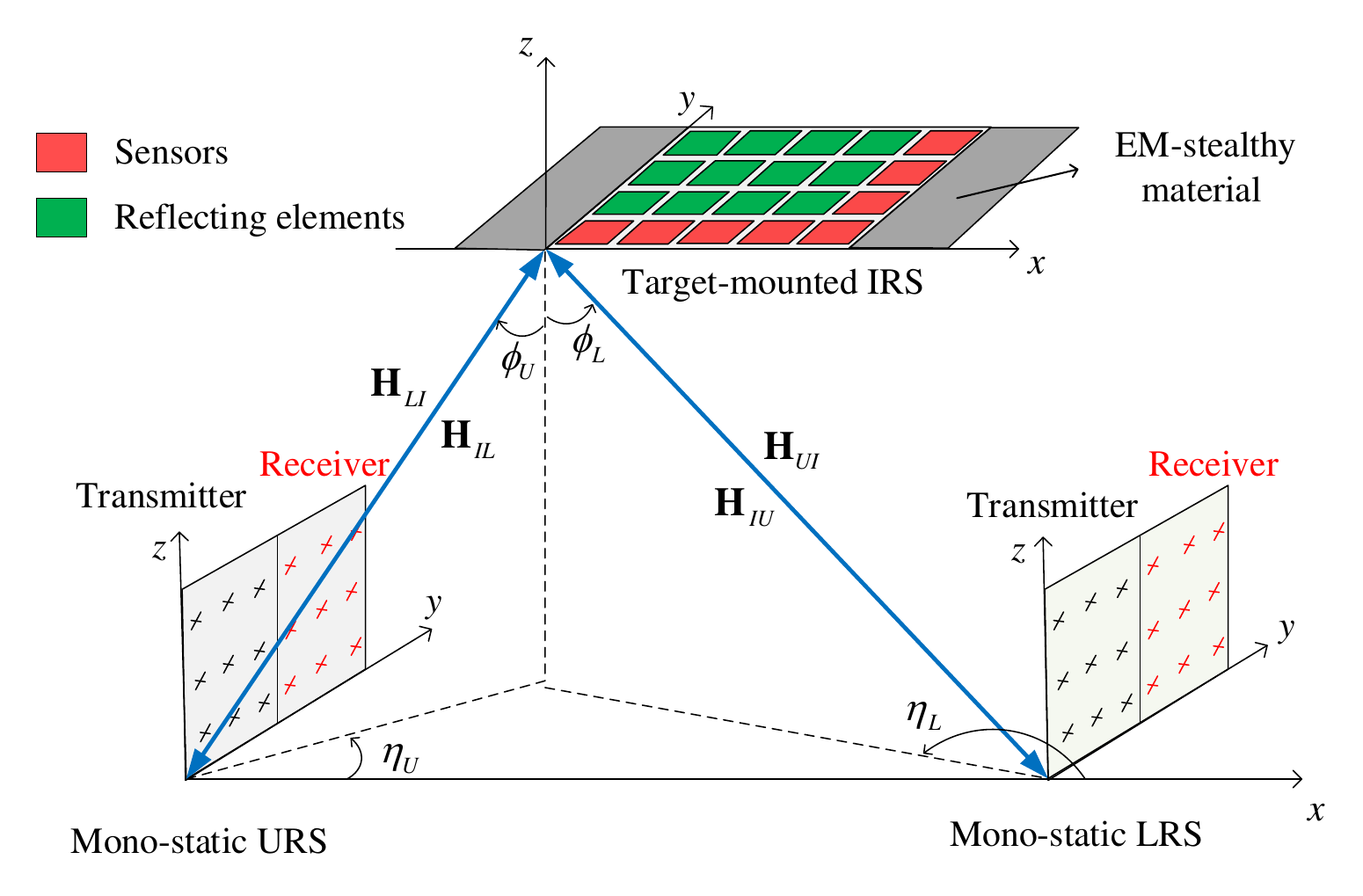}
\caption{Channel model.}
\label{channel_model}
\end{figure}

As shown in Fig. \ref{channel_model}, we consider that the LRS/URS is equipped with a uniform planar array (UPA) parallel to the $y-z$ plane. The LRS comprises $M=M_{{y}}\times M_{{z}}$ transmit/receive antennas, with $M_{{y}}$ and $M_{{z}}$ denoting the numbers of transmit/receive antennas along the $y$- and $z$-axes, respectively. The URS consists of $D=D_{{y}}\times D_{{z}}$ transmit/receive antennas, with $D_{{y}}$ and $D_{{z}}$ being the corresponding numbers of antennas along the $y$- and $z$-axes. The IRS is assumed to be another UPA with $N=N_y\times N_x$ reflecting elements, where
$N_y$ and $N_x$ are the respective numbers of reflecting elements along the $y$- and $x$-axes, and a controller is attached to the IRS for performing signal processing and adjusting its reflection over time. At the IRS, each reflecting element re-scatters the received signal with an independently tunable reflection coefficient. Specifically, we denote the IRS reflection coefficients as $\boldsymbol{\theta}=[\beta_1 e^{j\omega_1}, \cdots, \beta_N e^{j\omega_N}]^T$, where each reflection coefficient comprises a phase shift $\omega_n \in [0,2\pi)$ and an on/off amplitude parameter $\beta_n \in \{0,1\}, n=1,2,\cdots,N$. Note that $\beta_n=1$ means that the $n$-th reflecting element fully reflects the incident signal without amplitude change, while $\beta_n=0$ implies that the incident signal is fully absorbed by this reflecting element (similar to EM wave absorbing material). Moreover, to enable its sensing function, we assume that the IRS is equipped with $L=L_y+L_x-1$ sensors (for receiving signals only) with $L_y$ and $L_x$ denoting the number of sensors along the $y$- and $x$-axes, respectively (see Fig. \ref{channel_model}), which is also termed as ``semi-passive" IRS \cite{qing}.

Due to the high altitude of aerial targets, we assume that the channels between the LRS/URS and the target-mounted IRS follow the LoS model (otherwise, reliable sensing cannot be achieved). As the URS/LRS is assumed to be
mono-static radar where the transmit and receive arrays are placed at the same location, the AoD from its transmitter to the IRS is equivalent to the AoA from the IRS to its receiver. As depicted in Fig. \ref{channel_model}, we select the bottom-left elements of the UPAs at the LRS/URS transmitter,  LRS/URS receiver, and IRS as their respective reference elements for modeling the LoS channels between them (to be specified later). Furthermore, let $\phi_{U}$ and $\eta_{U}$ denote the URS's elevation and azimuth AoD/AoA with respect to (w.r.t.) the IRS, and ${\phi}_{{L}}$ and ${\eta}_{{L}}$ denote the elevation and azimuth AoD/AoA at the LRS w.r.t. the IRS, respectively. Note that due to the far-field propagation between the LRS/URS and IRS (or target), the above angles apply to all antenna/reflecting/sensor elements on their corresponding UPAs.

Let $\mathbf{d}(B,\zeta)$ denote the one-dimensional (1D) steering vector function of a given 1D array, which is defined as
\begin{align}\label{anz}
\mathbf{d}(B,\zeta)&=\left[1,e^{j\frac{2\pi {d}}{\lambda}\zeta},\cdots,e^{j\frac{2\pi {d}}{\lambda}(B-1)\zeta}\right]^T,
\end{align}
where $B$ denotes the array size, $\zeta$ denotes the steering
angle, $\lambda$ denotes the signal wavelength, and ${d}$ denotes the
distance between any two adjacent antenna/reflecting/sensor elements. Accordingly, assuming that the IRS's UPA is parallel to the $x$-$y$ plane, the two-dimensional (2D) steering vectors at the IRS w.r.t. LRS for the incident and reflected signals can be expressed as
\begin{align}\label{abb1}
\mathbf{a}_{{l}}({\phi}_{{L}},{\eta}_{{L}})&=\mathbf{d}(N_x,\zeta_l^a)\otimes \mathbf{d}(N_y,\zeta_l^e),\\
\mathbf{a}_{{l}}(\pi-{\phi}_{{L}},\pi+{\eta}_{{L}})&=\mathbf{d}(N_x,\zeta_l^{\bar{a}})\otimes \mathbf{d}(N_y,\zeta_l^{\bar{e}}),
\end{align}
respectively, where $\zeta_l^a\triangleq \sin({\phi}_{{L}})\cos({\eta}_{{L}})$,
$\zeta_l^e\triangleq \sin({\phi}_{{L}})\sin({\eta}_{{L}})$, $\zeta_l^{\bar{a}}\triangleq \sin(\pi-{\phi}_{{L}})\cos(\pi+{\eta}_{{L}})\triangleq -\sin({\phi}_{{L}})\cos({\eta}_{{L}})$,
and $\zeta_l^{\bar{e}}\triangleq \sin(\pi-{\phi}_{{L}})\sin(\pi+{\eta}_{{L}})\triangleq -\sin({\phi}_{{L}})\sin({\eta}_{{L}})$.
Similarly, the 2D steering vectors for the incident and reflected signals at the IRS w.r.t. URS are respectively given by
\begin{align}\label{abbb1}
\mathbf{a}_{{u}}({\phi}_{{U}},{\eta}_{{U}})&=\mathbf{d}(N_x,\zeta_u^a)\otimes \mathbf{d}(N_y,\zeta_u^e),\\
\mathbf{a}_{{u}}(\pi-{\phi}_{{U}},\pi+{\eta}_{{U}})&=\mathbf{d}(N_x,\zeta_u^{\bar{a}})\otimes \mathbf{d}(N_y,\zeta_u^{\bar{e}}),
\end{align}
where $\zeta_u^{a}\triangleq \sin({\phi}_{{U}})\cos({\eta}_{{U}})$,
$\zeta_u^{e}\triangleq \sin({\phi}_{{U}})\sin({\eta}_{{U}})$,
$\zeta_u^{\bar{a}}\triangleq \sin(\pi-{\phi}_{{U}})\cos(\pi+{\eta}_{{U}})\triangleq -\sin({\phi}_{{U}})\cos({\eta}_{{U}})$, and $\zeta_u^{\bar{e}}\triangleq \sin(\pi-{\phi}_{{U}})\sin(\pi-{\eta}_{{U}})\triangleq -\sin({\phi}_{{U}})\sin({\eta}_{{U}})$.

In addition, for the LRS's UPA assumed to be parallel to the $y$-$z$ plane, its transmit and receive 2D steering vectors w.r.t. the IRS are respectively given by
\begin{align}\label{ab1}
\mathbf{b}({\phi}_{{L}},{\eta}_{{L}})&=\mathbf{d}(M_{{y}},\varphi_t^a)\otimes \mathbf{d}(M_{{z}},\varphi_t^e),\\
\mathbf{b}(\pi-{\phi}_{{L}},\pi+{\eta}_{{L}})&=
\mathbf{d}(M_{{y}},\varphi_r^{a})\otimes \mathbf{d}(M_{{z}},\varphi_r^{e}),
\end{align}
where $\varphi_t^a\triangleq\sin({\phi}_{{L}})\sin({\eta}_{{L}})$, $\varphi_t^e\triangleq\cos({\phi}_{{L}})$, $\varphi_r^{a}\triangleq\sin(\pi-{\phi}_{{L}})\sin(\pi+{\eta}_{{L}})
\triangleq-\sin({\phi}_{{L}})\sin({\eta}_{{L}})$, and $\varphi_r^{e}\triangleq\cos(\pi-{\phi}_{{L}})\triangleq-\cos({\phi}_{{L}})$.
Similarly, assuming that the URS is parallel to the $y$-$z$ plane, its 2D transmit and receive steering vectors are respectively expressed as
\begin{align}\label{nu}
\mathbf{c}({\phi}_{{U}},{\eta}_{{U}})&=\mathbf{d}(D_{{y}},\upsilon_t^a)\otimes \mathbf{d}(D_{{z}},\upsilon_t^e),\\
\mathbf{c}(\pi-{\phi}_{{U}},\pi+{\eta}_{{U}})&=\mathbf{d}(D_{{y}},\upsilon_r^a)\otimes \mathbf{d}(D_{{z}},\upsilon_r^e),
\end{align}
where $\upsilon_t^a\triangleq\sin({\phi}_{{U}})\sin({\eta}_{{U}})$, $\upsilon_t^e\triangleq\cos({\phi}_{{U}})$, $\upsilon_r^a\triangleq\sin(\pi-{\phi}_{{U}})\sin(\pi+{\eta}_{{U}})
\triangleq-\sin({\phi}_{{U}})\sin({\eta}_{{U}})$, and $\upsilon_r^e\triangleq\cos(\pi-{\phi}_{{U}})
\triangleq-\cos({\phi}_{{U}})$.

As a result, the IRS$\rightarrow$LRS channel $\mathbf{H}_{LI}\in \mathbb{C}^{M\times N}$, LRS$\rightarrow$IRS channel $\mathbf{H}_{IL}\in \mathbb{C}^{N\times{M}}$, IRS$\rightarrow$URS channel $\mathbf{H}_{UI}\in \mathbb{C}^{D\times{N}}$, and URS$\rightarrow$IRS channel $\mathbf{H}_{IU}\in \mathbb{C}^{N\times{D}}$ can be written as
\begin{align}\label{LL}
 \mathbf{H}_{LI}&=\alpha_l\mathbf{b}(\pi-{\phi}_{{L}},\pi+{\eta}_{{L}})\mathbf{a}_{{l}}^H(\pi-{\phi}_{{L}},\pi+{\eta}_{{L}}), \\
\mathbf{H}_{IL}&=\alpha_l\mathbf{a}_{{l}}({\phi}_{{L}},{\eta}_{{L}})\mathbf{b}^H({\phi}_{{L}},{\eta}_{{L}}),\\
\mathbf{H}_{UI}&=\alpha_u\mathbf{c}(\pi-{\phi}_{{U}},\pi+{\eta}_{{U}})\mathbf{a}_{{u}}^H(\pi-{\phi}_{{U}},\pi+{\eta}_{{U}}),\\
\mathbf{H}_{IU}&=\alpha_u\mathbf{a}_{{u}}({\phi}_{{U}},{\eta}_{{U}})\mathbf{c}^H({\phi}_{{U}},{\eta}_{{U}}),
\end{align}
respectively, where $\alpha_l=e^{j\nu_l}\bar{\alpha}_l$ and $\alpha_u=e^{j\nu_u}\bar{\alpha}_u$ with $\bar{\alpha}_l$ and $\bar{\alpha}_u$ denoting the real-valued path gain of the LRS-IRS and URS-IRS LoS channels, respectively, and $\nu_l=\frac{2\pi d_{LI}}{\lambda}$ and $\nu_u=\frac{2\pi d_{UI}}{\lambda}$ being the reference phase of the LRS-IRS and URS-IRS channels, respectively. In the above, $d_{LI}$ and $d_{UI}$ represent the distance between the LRS and target and that between the URS and target, respectively, w.r.t. their reference elements. To characterize the effect of small variations of the LRS-/URS- target distance  (i.e., $d_{LI}$ or $d_{UI}$) due to target position local perturbation on the signal phases $\nu_l$ and $\nu_u$, we model $\nu_l$ and $\nu_u$ as independent and  uniformly distributed random variables in $[0, 2\pi)$ \cite{liang}.
\subsection{Signal Model}
\begin{figure*}[t!]
\setlength{\abovecaptionskip}{-0.cm}
\setlength{\belowcaptionskip}{0.cm}
  \centering
\includegraphics[width=0.67\textwidth]{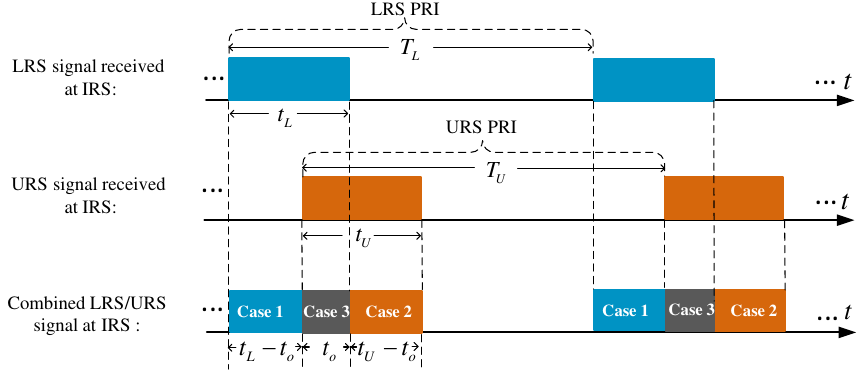}
\caption{Illustration of received LRS/URS signals at IRS.}
\label{pulse}
\end{figure*}
As shown in Fig. \ref{pulse}, we assume that the LRS transmits one coherent burst of $K_L$ non-consecutive radar pulses with a constant pulse repetition interval (PRI), denoted as $T_L$. The duration over which all these signals are reflected by the target and received by the LRS is called the coherent-processing interval (CPI), denoted by $T_{\mathrm{CPI}}$, which is equal to $K_L\times T_L$. The pulse durations of LRS and URS are denoted as $t_L$ and $t_U$, respectively, with $t_L<T_L$ and $t_U<T_U$, where the PRI of the URS is denoted as $T_U$.
For simplicity, we assume that $T_L=T_U=T$ in this paper\footnote{The proposed design in this paper can be extended to the general case of $T_L\neq T_U$, which is omitted due to space limitations.}. Furthermore, we assume that the target location as well as the PRIs and pulse durations of both LRS and URS remain unchanged during each LRS CPI, but they may change from one CPI to another.
In addition, we express the radar pulse waveforms of LRS and URS during  each PRI respectively as
\begin{align}\label{wl}
x(t)=\begin{cases}
\sqrt{P_{L}}p(t), & 0\leq t\leq t_L \\
0, &  t_L< t\leq T \\
\end{cases}
\end{align}
and
\begin{align}\label{wl2}
s(t)=\begin{cases}
\sqrt{P_{U}}q(t), & 0\leq t\leq t_U \\
0, &  t_U< t\leq T \\
\end{cases}
\end{align}
where $P_{L}$ and $P_{U}$ represent the transmit signal power of LRS and URS, respectively; $p(t)$ and $q(t)$ are the corresponding
radar pulses with normalized power, i.e., $\frac{1}{t_L}\int_{0}^{t_L}|p(t)|^2dt=1$ and $\frac{1}{t_U}\int_{0}^{t_U}|q(t)|^2dt=1$.

As shown in Fig. \ref{pulse}, the
signal received at the IRS during each LRS CPI may fall into the following three cases depending on whether the received signals from LRS and URS are overlapped or not, i.e.,
Case 1 (LRS-signal only case): the received signal is from LRS only; Case 2 (URS-signal only case): the received signal is from URS only; and Case 3 (overlapped LRS and URS signal): the received signal consists of overlapped signals from both LRS and URS. For convenience, we use $t_o$ to denote the overlapped signal duration in Case 3, with $0\leq t_o\leq \min\{t_L, t_U\}$. Note that if $t_o=0$, then the signals from LRS and URS are completely non-overlapped at the IRS.
Without loss of generality, assuming that the first LRS signal arrives at IRS before that of URS during each LRS CPI, we present the signal models for the above three cases in the following, respectively.

First, for Case 1 (i.e., $0\leq t \leq t_L-t_o $) with only the LRS signal reflected by the IRS, the signals received by the LRS receiver through the link \ding{172}$\rightarrow$\ding{173} in Fig. \ref{sys_model} (i.e., LRS$\rightarrow $target/IRS$\rightarrow$LRS) and the URS receiver via the link \ding{172}$\rightarrow$\ding{175} in Fig. \ref{sys_model} (i.e., LRS$\rightarrow$target/IRS$\rightarrow$URS) at time $t$ can be respectively expressed as
\begin{align}
y_{LL}(t)&=\mathbf{w}_{{L}}^T\mathbf{H}_{LI}
\text{diag}(\boldsymbol{\theta})\mathbf{H}_{IL}\mathbf{w}_{{L}}x(t)+z_{L}(t),\label{lrs1}\\
y_{LU}(t)&=\mathbf{w}_{{U}}^T\mathbf{H}_{UI}\text{diag}(\boldsymbol{\theta})\mathbf{H}_{IL}
\mathbf{w}_{{L}}x(t)+z_{U}(t),\label{lrs11}
\end{align}
where $z_{L}(t)$ and $z_{U}(t)$ denote the additive white Gaussian noise (AWGN) with zero mean and average power $\sigma_{L}^2$ and $\sigma_{U}^2$, respectively, $\mathbf{w}_{{L}}\in \mathbb{C}^{M \times 1}$ and $\mathbf{w}_{{U}} \in \mathbb{C}^{D \times 1}$ are the transmit beamformers at LRS and URS, respectively, and $\mathbf{w}_{{L}}^T\in \mathbb{C}^{1 \times M}$ and $\mathbf{w}_{{U}}^T\in \mathbb{C}^{1 \times D}$ denote their matching receive beamformers at LRS and URS, respectively.

Second, for Case 2 (i.e., $t_L < t \leq t_L+t_U-t_o$) with only the URS signal reflected by the IRS, the signals
received by the LRS receiver via the link \ding{174}$\rightarrow$\ding{173} in Fig. \ref{sys_model} (i.e., URS$\rightarrow $target/IRS$\rightarrow$LRS) and the URS receiver through the link \ding{174}$\rightarrow$\ding{175} in Fig. \ref{sys_model} (i.e., URS$\rightarrow $target/IRS$\rightarrow$URS)  are respectively given by
\begin{align}
y_{UL}(t)&=\mathbf{w}_{{L}}^T\mathbf{H}_{LI}\text{diag}(\boldsymbol{\theta})\mathbf{H}_{IU}
\mathbf{w}_{{U}}s(t)+z_{L}(t),\label{lrs22}\\
y_{UU}(t)&=\mathbf{w}_{{U}}^T\mathbf{H}_{UI}
\text{diag}(\boldsymbol{\theta})\mathbf{H}_{IU}\mathbf{w}_{{U}}s(t)+z_{U}(t).\label{lrs2}
\end{align}

Third, for Case 3 (i.e., $t_L-t_o<t \leq t_L$) with the overlapped signal reflected by the IRS, the signals received by the LRS receiver through both links \ding{172}$\rightarrow$\ding{173} and \ding{174}$\rightarrow$\ding{173} in Fig. \ref{sys_model} and the URS
receiver via both links \ding{172}$\rightarrow$\ding{175} and \ding{174}$\rightarrow$\ding{175} in Fig. \ref{sys_model} are respectively expressed as
\begin{align}
y_{OL}(t)&=\mathbf{w}_{{L}}^T\mathbf{H}_{LI}
\text{diag}(\boldsymbol{\theta})\mathbf{H}_{IL}\mathbf{w}_{{L}}x(t)\nonumber\\
&+\mathbf{w}_{{L}}^T\mathbf{H}_{LI}\text{diag}(\boldsymbol{\theta})\mathbf{H}_{IU}
\mathbf{w}_{{U}}s(t)+z_{L}(t),\label{lrs3}\\
y_{OU}(t)&=\mathbf{w}_{{U}}^T\mathbf{H}_{UI}
\text{diag}(\boldsymbol{\theta})\mathbf{H}_{IU}\mathbf{w}_{{U}}s(t)\nonumber\\
&+\mathbf{w}_{{U}}^T\mathbf{H}_{UI}\text{diag}(\boldsymbol{\theta})\mathbf{H}_{IL}
\mathbf{w}_{{L}}x(t)+z_{U}(t).\label{lrs33}
\end{align}

In practice, the performance of target detection/estimation (e.g., detecting the presence of a target or estimating a target's AoA) improves with the increase of received signal power \cite{power}. This is intuitively expected, since larger signal power results in higher signal-to-noise ratio (SNR) of the received echo signal, thus leading to lower detection/estimation error. Therefore, we use the received signal powers at the LRS/URS as the performance metric for their target detection/estimation, and derive them for the above three cases, respectively, as follows.

First, according to \eqref{lrs1} and \eqref{lrs11}, the received signal powers at LRS and URS in Case 1 are respectively given by
\begin{align}\label{qll}
{Q}_{LL}&={|\mathbf{w}_{{L}}^T\mathbf{H}_{LI}
\text{diag}(\boldsymbol{\theta})\mathbf{H}_{IL}\mathbf{w}_{{L}}x(t)|^2}\nonumber\\
&={|\bar{\alpha}_l\mathbf{w}_{{L}}^T\mathbf{b}(\pi-{\phi}_{{L}},\pi+{\eta}_{{L}})
{\mathbf{u}}^H
\boldsymbol{\theta}\bar{\alpha}_l
\mathbf{b}^H({\phi}_{{L}},{\eta}_{{L}})\mathbf{w}_{{L}}x(t)|^2}\nonumber\\
&={\frac{Q_{LS}^2}{P_L}\times|{\mathbf{u}}^H
\boldsymbol{\theta}|^2},
\end{align}
and
\begin{align}\label{qlu}
{Q}_{LU}&={|\mathbf{w}_{{U}}^T\mathbf{H}_{UI}\text{diag}(\boldsymbol{\theta})\mathbf{H}_{IL}
\mathbf{w}_{{L}}x(t)|^2}\nonumber\\
&={|\bar{\alpha}_u\mathbf{w}_{{U}}^T\mathbf{c}(\pi-{\phi}_{{U}},\pi+{\eta}_{{U}}){\mathbf{v}}^H
\boldsymbol{\theta}\bar{\alpha}_l\mathbf{b}^H({\phi}_{{L}},{\eta}_{{L}})
\mathbf{w}_{{L}}x(t)|^2}\nonumber\\
&={\frac{Q_{LS}Q_{US}}{P_U}\times|{\mathbf{v}}^H
\boldsymbol{\theta}|^2},
\end{align}
with
\begin{align}
{\mathbf{u}}&=\mathbf{a}_l(\pi-{\phi}_{{L}},\pi+{\eta}_{{L}})\odot \mathbf{a}_l^*({\phi}_{{L}},{\eta}_{{L}})\nonumber\\
&\triangleq \mathbf{d}(N_x,\zeta_l^{\bar{a}}-\zeta_l^{a}) \otimes \mathbf{d}(N_y,\zeta_l^{\bar{e}}-\zeta_l^{e})\in \mathbb{C}^{N\times 1},\label{uo}\\
{\mathbf{v}}&=\mathbf{a}_u(\pi-{\phi}_{{U}},\pi+{\eta}_{{U}})\odot \mathbf{a}_l({\phi}_{{L}},{\eta}_{{L}}) \nonumber\\ &\triangleq \mathbf{d}(N_x,\zeta_u^{\bar{a}}-\zeta_l^{a}) \otimes \mathbf{d}(N_y,\zeta_u^{\bar{e}}-\zeta_l^{e})\in \mathbb{C}^{N\times 1}, \label{vo}\\
Q_{LS}&=|\bar{\alpha}_l
\mathbf{b}^H({\phi}_{{L}},{\eta}_{{L}})\mathbf{w}_{{L}}x(t)|^2,\label{PLI}\\
Q_{US}&=|\bar{\alpha}_u\mathbf{w}_{{U}}^T\mathbf{c}(\pi-{\phi}_{{U}},\pi+{\eta}_{{U}})s(t)|^2,\label{PUI}
\end{align}
where $|{\mathbf{u}}^H \boldsymbol{\theta}|^2$ and $|{\mathbf{v}}^H \boldsymbol{\theta}|^2$ depend on the IRS reflection, i.e., $\boldsymbol{\theta}$, and $Q_{LS}$ and $Q_{US}$ represent the received signal powers at IRS from LRS and URS, respectively.

Similarly, based on \eqref{lrs22} and \eqref{lrs2}, the received signal powers at LRS and URS in Case 2 are respectively
\begin{align}\label{qul}
{Q}_{UL}&={|\mathbf{w}_{{L}}^T\mathbf{H}_{LI}\text{diag}(\boldsymbol{\theta})\mathbf{H}_{IU}
\mathbf{w}_{{U}}s(t)|^2}\nonumber\\
&={|\bar{\alpha}_l\mathbf{w}_{{L}}^T\mathbf{b}(\pi-{\phi}_{{L}},\pi+{\eta}_{{L}}){\mathbf{r}}^H
\boldsymbol{\theta}\bar{\alpha}_u\mathbf{c}^H
({\phi}_{{U}},{\eta}_{{U}})
\mathbf{w}_{{U}}s(t)|^2}\nonumber\\
&={\frac{Q_{LS}Q_{US}}{P_L}\times|{\mathbf{r}}^H
\boldsymbol{\theta}|^2},
\end{align}
and
\begin{align}\label{quu}
{Q}_{UU}&={|\mathbf{w}_{{U}}^T\mathbf{H}_{UI}
\text{diag}(\boldsymbol{\theta})\mathbf{H}_{IU}\mathbf{w}_{{U}}s(t)|^2}\nonumber\\
&={|\bar{\alpha}_u\mathbf{w}_{{U}}^T\mathbf{c}(\pi-{\phi}_{{U}},\pi+{\eta}_{{U}}){\mathbf{g}}^H
\boldsymbol{\theta}\bar{\alpha}_u
\mathbf{c}^H({\phi}_{{U}},{\eta}_{{U}})\mathbf{w}_{{U}}s(t)|^2}\nonumber\\
&={\frac{Q_{US}^2}{P_U}\times|{\mathbf{g}}^H
\boldsymbol{\theta}|^2},
\end{align}
with
\begin{align}
{\mathbf{r}}&=\mathbf{a}_l(\pi-{\phi}_{{L}},\pi+{\eta}_{{L}})\odot \mathbf{a}_u^*({\phi}_{{U}},{\eta}_{{U}})\nonumber\\
&\triangleq \mathbf{d}(N_x,\zeta_l^{\bar{a}}-\zeta_u^{a}) \otimes \mathbf{d}(N_y,\zeta_l^{\bar{e}}-\zeta_u^{e})\in \mathbb{C}^{N\times 1},\label{ro}\\
{\mathbf{g}}&=\mathbf{a}_u(\pi-{\phi}_{{U}},\pi+{\eta}_{{U}})\odot \mathbf{a}_u^*({\phi}_{{U}},{\eta}_{{U}})\nonumber\\
&\triangleq \mathbf{d}(N_x,\zeta_u^{\bar{a}}-\zeta_u^{a}) \otimes \mathbf{d}(N_y,\zeta_u^{\bar{e}}-\zeta_u^{e})\in \mathbb{C}^{N\times 1},\label{go}
\end{align}
where $|{\mathbf{r}}^H \boldsymbol{\theta}|^2$ and $|{\mathbf{g}}^H \boldsymbol{\theta}|^2$ depend on the IRS reflection, i.e., $\boldsymbol{\theta}$.

Furthermore, from \eqref{lrs3} and \eqref{lrs33}, the average received signal powers at LRS and URS in Case 3 are respectively obtained as
\begin{align}
{Q}_{OL}&=\mathbb{E}\left[|\mathbf{w}_{{L}}^T\mathbf{H}_{LI}
\text{diag}(\boldsymbol{\theta})\mathbf{H}_{IL}\mathbf{w}_{{L}}x(t)\right.\nonumber\\
&\left.+\mathbf{w}_{{L}}^T\mathbf{H}_{LI}\text{diag}(\boldsymbol{\theta})\mathbf{H}_{IU}
\mathbf{w}_{{U}}s(t)|^2\right]\\
&={|\mathbf{f}|^2}
+{|\bar{\mathbf{f}}|^2} +\mathbb{E}[e^{j\nu_l}]\mathbb{E}[e^{-j\nu_u}]\mathbf{f}\bar{\mathbf{f}}^*
\nonumber\\
&+\mathbb{E}[e^{-j\nu_l}]\mathbb{E}[e^{j\nu_u}]\mathbf{f}^*\bar{\mathbf{f}}
\label{oll}\\
&={\frac{Q_{LS}^2}{P_L}\times|{\mathbf{u}}^H
\boldsymbol{\theta}|^2}+{\frac{Q_{LS}Q_{US}}{P_L}\times|{\mathbf{r}}^H
\boldsymbol{\theta}|^2},\label{qOL}
\end{align}
and
\begin{align}
{Q}_{OU}&=\mathbb{E}\left[|\mathbf{w}_{{U}}^T\mathbf{H}_{UI}\text{diag}(\boldsymbol{\theta})\mathbf{H}_{IL}
\mathbf{w}_{{L}}x(t)\right.\nonumber\\
&\left.+\mathbf{w}_{{U}}^T\mathbf{H}_{UI}
\text{diag}(\boldsymbol{\theta})\mathbf{H}_{IU}\mathbf{w}_{{U}}s(t)|^2\right]\\
&={|\mathbf{e}|^2}
+{|\bar{\mathbf{e}}|^2} +\mathbb{E}[e^{j\nu_l}]\mathbb{E}[e^{-j\nu_u}]\mathbf{e}\bar{\mathbf{e}}^*
\nonumber\\
&+\mathbb{E}[e^{-j\nu_l}]\mathbb{E}[e^{j\nu_u}]\mathbf{e}^*\bar{\mathbf{e}} \label{3oll}\\
&={\frac{Q_{LS}Q_{US}}{P_U}\times|{\mathbf{v}}^H
\boldsymbol{\theta}|^2}+{\frac{Q_{US}^2}{P_U}\times|{\mathbf{g}}^H
\boldsymbol{\theta}|^2},\label{3qOL}
\end{align}
where $\mathbf{f}=e^{2j\nu_l}\bar{\alpha}_l^2\mathbf{w}_{{L}}^T
\mathbf{b}(\pi-{\phi}_{{L}},\pi+{\eta}_{{L}})
{\mathbf{u}}^H
\boldsymbol{\theta}
\mathbf{b}^H({\phi}_{{L}},{\eta}_{{L}})\mathbf{w}_{{L}}x(t)$, $\bar{\mathbf{f}}=e^{j\nu_l}e^{j\nu_u}\bar{\alpha}_l\bar{\alpha}_u\mathbf{w}_{{L}}^T\mathbf{b}
(\pi-{\phi}_{{L}},\pi+{\eta}_{{L}}){\mathbf{r}}^H
\boldsymbol{\theta} \mathbf{c}^H
({\phi}_{{U}},{\eta}_{{U}})
\mathbf{w}_{{U}}s(t)$, $\mathbf{e}=e^{j\nu_l}e^{j\nu_u}\bar{\alpha}_u\bar{\alpha}_l\mathbf{w}_{{U}}^T
\mathbf{c}(\pi-{\phi}_{{U}},\pi+{\eta}_{{U}}){\mathbf{v}}^H
\boldsymbol{\theta}\mathbf{b}^H({\phi}_{{L}},{\eta}_{{L}})
\mathbf{w}_{{L}}x(t)$, and $\bar{\mathbf{e}}=e^{2j\nu_u}\bar{\alpha}_u^2\mathbf{w}_{{U}}^T
\mathbf{c}(\pi-{\phi}_{{U}},\pi+{\eta}_{{U}}){\mathbf{g}}^H
\boldsymbol{\theta}
\mathbf{c}^H({\phi}_{{U}},{\eta}_{{U}})\mathbf{w}_{{U}}s(t)$.
In the above, the expectations are taken w.r.t. the independent random phases  $\nu_l$ and $\nu_u$, while \eqref{oll} and \eqref{3oll} hold since $\mathbb{E}[e^{\pm j\nu_l}]=0$ and $\mathbb{E}[e^{\pm j\nu_u}]=0$.

\section{Proposed Protocol for Target-Mounted IRS}
\begin{figure*}[t!]
\setlength{\abovecaptionskip}{-0.cm}
\setlength{\belowcaptionskip}{0.cm}
  \centering
\includegraphics [width=0.61\textwidth]{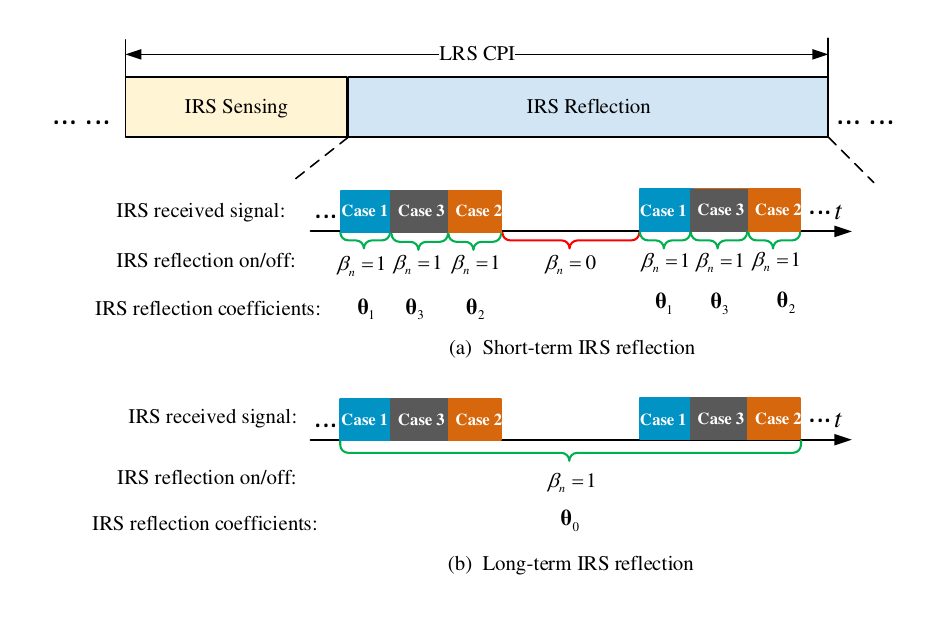}
\caption{Proposed protocol for secure wireless sensing with the target-mounted IRS.}
\label{waveformf}
\end{figure*}
In this section, we propose a practical protocol for the operation of the target-mounted IRS for secure wireless sensing, as illustrated in Fig. \ref{waveformf}. The protocol consists of two steps in each LRS CPI.
In the first step (Step I), IRS reflecting elements are switched off and IRS sensors estimate the required LRS/URS channel and waveform parameters. In the second step (Step II), based on the estimated information, IRS controller first designs the phase shifts of IRS reflecting elements, and then they are switched on/off with their phase shifts set accordingly. In the following, we elaborate the above two steps in detail.

In Step I, with all IRS reflecting elements switched off, i.e., $\beta_n=0, \forall n=1,...,N$, IRS uses its sensors to receive the radar signals from LRS and URS for estimating their respective AoAs, i.e., $\{\phi_L,\eta_L\}$ and $\{\phi_U,\eta_U\}$, which are required for the IRS reflection design in Step II. Specifically, these AoAs can be estimated by applying high-resolution angle estimation methods such as multiple signal classification (MUSIC)\cite{music}. In addition, the LRS/URS PRI $T$, their pulse durations, i.e., $t_{L}$ and $t_{U}$, and overlapped signal duration $t_o$ are also needed for the subsequent IRS reflection design, which, if not a priori known, can be estimated based on e.g, the deinterleaving approach \cite{dein,ide}. Specifically, the autocorrelation-based method and specialized finite impulse response filter can be utilized to extract different PRIs \cite{dein} and pulse durations \cite{filter, survey}, respectively, and then these estimated parameters are compared with their known values stored for LRS to identify those corresponding to URS, which also helps identify the previously estimated AoAs belonging to LRS or URS. Finally, the received signal powers at IRS from LRS/URS, i.e., $Q_{LS}$ and $Q_{US}$, are also required for the design of IRS reflection, which can be measured from the received signals at IRS sensors after resolving the
LRS/URS AoAs by e.g., the power gain estimation method \cite{mea}, even when there is signal overlap at the IRS from both LRS and URS (i.e., Case 3)\cite{com}.

With the above estimated parameters, IRS controller then designs the phase shifts of its reflecting elements in Step II for simultaneously boosting the received signal at the LRS receiver and suppressing that at the URS receiver (see Section IV for details). In particular, we consider two IRS reflection design approaches, namely short-term or long-term IRS reflection, which offer different trade-offs between performance and complexity. Specifically,
in the short-term IRS reflection case, as shown in Fig. \ref{waveformf}(a), the IRS reflecting elements are
switched on (i.e., $\beta_n=1, \forall n=1,...N$) in Step II during Cases 1, 2, and 3, with the corresponding reflection phase shifts given by $\boldsymbol{\theta}_1$, $\boldsymbol{\theta}_2$, and $\boldsymbol{\theta}_3$, respectively; otherwise, the IRS reflecting elements are switched off (i.e., $\beta_n=0, \forall n=1,...N$) to save power. In contrast, with long-term IRS reflection, as shown in Fig. \ref{waveformf}(b), the IRS reflecting elements are switched on (i.e., $\beta_n=1, \forall n=1,...N$) during the entire Step II, with fixed IRS reflection phase shifts given by $\boldsymbol{\theta}_0$.

Comparing the short-term and long-term IRS reflections, we can observe that
short-term reflection offers more flexible IRS passive beamforming design for different cases and thus is expected to achieve better performance (in terms of enhanced/suppressed signal power at LRS/URS) than long-term reflection.
However, it also incurs higher computational and implementation complexities due to more design variables, and more frequent IRS phase shift adjustment.

\section{IRS Reflection Design}
In this section, we consider IRS reflection design in Step II of the proposed protocol for short-term and long-term IRS reflections, respectively.

\subsection{Short-Term IRS Reflection Design}
For short-term IRS reflection, we need to optimize the IRS reflection separately for three cases within each LRS CPI, namely, LRS-signal only, URS-signal only, and overlapped LRS and URS signal. Specifically, for the LRS-signal-only case, we aim to maximize the power of the IRS-reflected LRS signal over the LRS$\rightarrow $IRS$\rightarrow$LRS link, while keeping that over the LRS$\rightarrow $IRS$\rightarrow$URS link below a certain level, by optimizing the IRS reflection phase shifts. Consequently, based on \eqref{qll} and \eqref{qlu}, the optimization problem can be formulated as
\begin{subequations}
\label{11}
\begin{align}
(\mathcal{P}_1): \mathop{\max}\limits_{\boldsymbol{\theta}_1}~&~ \underbrace{Q_{LS}^2\times |{\mathbf{u}}^H
\boldsymbol{\theta}_1|^2}_{\text{LRS}\rightarrow \text{IRS}\rightarrow \text{LRS}} \label{lo}
\\
\text {s.t.}~&~\underbrace{{\frac{Q_{LS}Q_{US}}{P_{U,\min}}\times|{\mathbf{v}}^H
\boldsymbol{\theta}_1|^2}}_{\text{LRS}\rightarrow \text{IRS}\rightarrow \text{URS}} \leq \gamma, \label{lo1}\\
~&~|{\theta}_{1,n}|= 1, n=1,\cdots,N,
\end{align}
\end{subequations}
where $P_{U,\min}$ denotes the minimum value of $P_{U}$ which is assumed to be known, and $\gamma$ denotes the maximum received signal power threshold at the URS, below which the URS cannot achieve its desired target detection/estimation performance.

Next, for the URS-signal-only case, our goal is to maximize
the power of IRS-reflected URS signal over the URS$\rightarrow $IRS$\rightarrow$LRS link for exploiting the URS's radar signal for the LRS's target detection, subject to
that over the URS$\rightarrow $IRS$\rightarrow$URS link being below the given threshold $\gamma$. According to \eqref{qul} and \eqref{quu}, the IRS reflection phase shifts can be optimized by solving the following problem,
\begin{subequations}
\label{22}
\begin{align}
(\mathcal{P}_2): \mathop{\max}\limits_{\boldsymbol{\theta}_2}~&~ \underbrace{Q_{LS}Q_{US}\times|{\mathbf{r}}^H
\boldsymbol{\theta}_2|^2}_{\text{URS}\rightarrow \text{IRS}\rightarrow \text{LRS}} \label{2lo1}
\\
\text { s.t. }~&~ \underbrace{{\frac{Q_{US}^2}{P_{U,\min}}\times|{\mathbf{g}}^H
\boldsymbol{\theta}_2|^2}}_{\text{URS}\rightarrow \text{IRS}\rightarrow \text{URS}}\leq \gamma,  \label{lo11}\\
~&~|{\theta}_{2,n}|= 1, n=1,\cdots,N.
\end{align}
\end{subequations}

Last, for the case of overlapped LRS and URS signal, we aim to maximize the average power of IRS-reflected overlapped LRS and URS signal at the LRS receiver, while keeping that at the URS receiver below $\gamma$. To achieve this, we can formulate the following IRS reflection phase shifts optimization problem based on \eqref{qOL} and \eqref{3qOL},
\begin{subequations}
\label{33}
\begin{align}
\!\!\!\!(\mathcal{P}_3): \mathop{\max}\limits_{\boldsymbol{\theta}_3}~&~
\underbrace{{{Q_{LS}^2}\times|{\mathbf{u}}^H
\boldsymbol{\theta}_3|^2}}_{\text{LRS}\rightarrow \text{IRS}\rightarrow \text{LRS}}+\underbrace{{{Q_{LS}Q_{US}}\times|{\mathbf{r}}^H
\boldsymbol{\theta}_3|^2}}_{\text{URS}\rightarrow \text{IRS}\rightarrow \text{LRS}}
\label{pj}\\
\text {s.t.}~&~\underbrace{{\frac{Q_{US}^2}{P_{U,\min}}\times|{\mathbf{g}}^H
\boldsymbol{\theta}_3|^2}}_{\text{URS}\rightarrow \text{IRS}\rightarrow \text{URS}}+\underbrace{{\frac{Q_{LS}Q_{US}}{P_{U,\min}}\times|{\mathbf{v}}^H
\boldsymbol{\theta}_3|^2}}_{\text{LRS}\rightarrow \text{IRS}\rightarrow \text{URS}}\leq \gamma,\label{pj1}\\
~&~|{\theta}_{3,n}|= 1, n=1,\cdots,N. \label{pj2}
\end{align}
\end{subequations}

{\bf{\emph{Remark 1}}}: The estimates of LRS/URS AoAs $\{\phi_L, \eta_L\}$ and $\{\phi_U, \eta_U\}$ in Step I of the proposed protocol are required for constructing the vectors $\mathbf{u}$, $\mathbf{v}$, $\mathbf{r}$, and $\mathbf{g}$ involved in problems $(\mathcal{P}_1)$ to $(\mathcal{P}_3)$. In addition, the estimates of received signal powers from LRS and URS at IRS, i.e., $Q_{LS}$ and $Q_{US}$ in Step I are also needed for formulating the objective functions and constraints in $(\mathcal{P}_1)$ to $(\mathcal{P}_3)$.

\subsection{Long-Term IRS Reflection Design}
The objective of long-term IRS reflection is to maximize the total energy of IRS-reflected LRS signal with duration $t_L$ and URS signal with duration $t_U$ at the LRS receiver, while keeping their overlapped signal peak power (in Case 3 by assuming the worst case of $t_o>0$) at the URS receiver below the given threshold $\gamma$. Similar to ($\mathcal{P}_3$), the IRS reflection phase shifts during Step II can be optimized by the following problem,
\begin{subequations}
\label{31}
\begin{align}
(\mathcal{P}_4): \mathop{\max}\limits_{\boldsymbol{\theta}_0}~&~
\underbrace{t_L{{Q_{LS}^2}\times|{\mathbf{u}}^H
\boldsymbol{\theta}_0|^2}}_{\text{LRS}\rightarrow \text{IRS}\rightarrow \text{LRS}}+\underbrace{t_U{{Q_{LS}Q_{US}}\times|{\mathbf{r}}^H
\boldsymbol{\theta}_0|^2}}_{\text{URS}\rightarrow \text{IRS}\rightarrow \text{LRS}}
\label{pqw}\\
\text {s.t.}~&~\underbrace{{\frac{Q_{US}^2}{P_{U,\min}}\times|{\mathbf{g}}^H
\boldsymbol{\theta}_0|^2}}_{\text{URS}\rightarrow \text{IRS}\rightarrow \text{URS}}+\underbrace{{\frac{Q_{LS}Q_{US}}{P_{U,\min}}\times|{\mathbf{v}}^H
\boldsymbol{\theta}_0|^2}}_{\text{LRS}\rightarrow \text{IRS}\rightarrow \text{URS}}\leq \gamma,\label{pqw1}\\
~&~|{\theta}_{0,n}|= 1, n=1,\cdots,N. \label{pqw2}
\end{align}
\end{subequations}

{\bf{\emph{Remark 2}}}: The estimation of AoA values $\{\phi_L, \eta_L\}$ and $\{\phi_U, \eta_U\}$, as well as received signal powers $Q_{LS}$ and $Q_{US}$ in Step I of the proposed protocol is required for formulating problem $(\mathcal{P}_4)$. In addition, the LRS and URS pulse durations, i.e., $t_L$ and $t_U$, respectively, are explicitly used for formulating the objective function of $(\mathcal{P}_4)$, which, if not a priori known, also need to be estimated in Step I of the proposed protocol.

Note that if the constraint \eqref{pqw1} in problem $(\mathcal{P}_4)$ is satisfied, then constraints \eqref{lo1} in problem $(\mathcal{P}_1)$, \eqref{lo11} in problem $(\mathcal{P}_2)$, and \eqref{pj1} in problem $(\mathcal{P}_3)$ are also satisfied. Thus, the optimal solution $\boldsymbol{\theta}_0^o$ to problem $(\mathcal{P}_4)$ is in general only a feasible solution to problems $(\mathcal{P}_1)$--$(\mathcal{P}_3)$. Consequently, we have the following inequality,
\begin{align}\label{eq}
&t_L\times{{Q_{LS}^2}|{\mathbf{u}}^H
\boldsymbol{\theta}_0^o|^2}+t_U\times{{Q_{LS}Q_{US}}|{\mathbf{r}}^H
\boldsymbol{\theta}_0^o|^2}\nonumber\\
&\leq (t_L-t_o)\times{{Q_{LS}^2}|{\mathbf{u}}^H
\boldsymbol{\theta}_1^o|^2}+(t_U-t_o)\times{{Q_{LS}Q_{US}}|{\mathbf{r}}^H
\boldsymbol{\theta}_2^o|^2} \nonumber\\
& + t_o\times ({{Q_{LS}^2}|{\mathbf{u}}^H
\boldsymbol{\theta}_3^o|^2}+{{Q_{LS}Q_{US}}|{\mathbf{r}}^H
\boldsymbol{\theta}_3^o|^2}),
\end{align}
where $\boldsymbol{\theta}_{1}^o$, $\boldsymbol{\theta}_{2}^o$, and $\boldsymbol{\theta}_{3}^o$ denote the optimal solutions to problems ($\mathcal{P}_1$)--($\mathcal{P}_3$), respectively. The above inequality indicates that the maximum LRS received signal energy with long-term IRS reflection is upper-bounded by that with short-term IRS reflection. In particular, when the LRS and URS waveforms have the same duration and they are completely overlapped, i.e., $t_o=t_L=t_U$, \eqref{eq} holds with equality since $t_L-t_o=0$, $t_U-t_o=0$, and  $\boldsymbol{\theta}_0^o=\boldsymbol{\theta}_3^o$. As a result, the long-term and short-term IRS reflections yield the same LRS received signal energy.

The above formulated optimization problems, namely $(\mathcal{P}_1)$--$(\mathcal{P}_4)$, share the same structure and can be all expressed by the following general problem,
\begin{subequations}
\begin{align}
(\mathcal{P}_5): \mathop{\max}\limits_{\boldsymbol{\theta}}~&~ \sum_{i=1}^{2}|\mathbf{q}_i^H
\boldsymbol{\theta}|^2 \label{OP01}\\
\text {s.t.}~&~\sum_{i=1}^{2}|{\mathbf{h}}_i^H
\boldsymbol{\theta}|^2\leq \gamma, \label{cc01}\\
~&~|{\theta}_{n}|= 1, n=1,\cdots,N, \label{cc03}
\end{align}
\end{subequations}
with $\mathbf{q}_1=\sqrt{{{Q_{LS}^2}}}\mathbf{u}$, $\mathbf{q}_2=\mathbf{0}$,
$\mathbf{h}_1=\sqrt{\frac{Q_{LS}Q_{US}}{P_{U,\min}}}\mathbf{v}$, and $\mathbf{h}_2=\mathbf{0}$ for problem $(\mathcal{P}_1)$;
$\mathbf{q}_1=\sqrt{{{Q_{LS}Q_{US}}}}\mathbf{r}$, $\mathbf{q}_2=\mathbf{0}$,
$\mathbf{h}_1=\sqrt{\frac{Q_{US}^2}{P_{U,\min}}}\mathbf{g}$, and $\mathbf{h}_2=\mathbf{0}$ for problem $(\mathcal{P}_2)$;
$\mathbf{q}_1=\sqrt{{{Q_{LS}^2}}}\mathbf{u}$, $\mathbf{q}_2=\sqrt{{{Q_{LS}Q_{US}}}}\mathbf{r}$,
$\mathbf{h}_1=\sqrt{\frac{Q_{US}^2}{P_{U,\min}}}\mathbf{g}$, and $\mathbf{h}_2=\sqrt{\frac{Q_{LS}Q_{US}}{P_{U,\min}}}\mathbf{v}$ for problem $(\mathcal{P}_3)$; and
$\mathbf{q}_1=\sqrt{t_L{{Q_{LS}^2}}}\mathbf{u}$, $\mathbf{q}_2=\sqrt{t_U{{Q_{LS}Q_{US}}}}\mathbf{r}$,
$\mathbf{h}_1=\sqrt{\frac{Q_{US}^2}{P_{U,\min}}}\mathbf{g}$, and $\mathbf{h}_2=\sqrt{\frac{Q_{LS}Q_{US}}{P_{U,\min}}}\mathbf{v}$ for problem $(\mathcal{P}_4)$.
As such, it suffices to solve problem $(\mathcal{P}_5)$. Note that  $(\mathcal{P}_5)$ can be reformulated such that the SDR method
can be used to solve it sub-optimally \cite{sdr}. However, the SDR-based method is computationally prohibitive, especially when the number of IRS reflecting elements is large. To reduce the computational complexity, we propose in the following an alternative PDD-based algorithm for solving $(\mathcal{P}_5)$ more efficiently.
\subsection{PDD-based Algorithm}
In this subsection, we present a computationally efficient algorithm to solve ($\mathcal{P}_5$).
It is noted that ($\mathcal{P}_5$) is a non-convex optimization problem because the objective in \eqref{OP01} is non-concave and the constraints in \eqref{cc03} are non-convex. Hence, we adopt the PDD method \cite{shi_pen}, which is based on the augmented Lagrangian and can separate the coupled optimization variables by penalizing the auxiliary equality constraints via dual decomposition. To apply the PDD method, we first decouple the constraints \eqref{cc01}
and \eqref{cc03} by introducing auxiliary variables $\boldsymbol{\vartheta}=[\vartheta_1,\vartheta_2,\cdots,\vartheta_N]^T\in \mathbb{C}^{N\times 1}$, which satisfy $\boldsymbol{\vartheta}=\boldsymbol{\theta}$. As a result, problem ($\mathcal{P}_5$) can be equivalently recast as
\begin{subequations}
\begin{align}
(\mathcal{P}_6): \mathop{\max}\limits_{\boldsymbol{\theta}}~&~ \sum_{i=1}^{2}|\mathbf{q}_i^H
\boldsymbol{\theta}|^2 \label{OP1}\\
\text {s.t.}~&~\sum_{i=1}^{2}|{\mathbf{h}}_i^H
\boldsymbol{\theta}|^2\leq \gamma, \label{cc1}\\
~&~\boldsymbol{\theta}=\boldsymbol{\vartheta},\label{cc2}\\
~&~|{\vartheta}_n|= 1, n=1,\cdots,N. \label{cc3}
\end{align}
\end{subequations}
Then, by penalizing the equality constraint in \eqref{cc2}, the dual optimization problem of ($\mathcal{P}_6$) can be written as
\begin{subequations}
\begin{align}
(\mathcal{P}_7): \mathop{\min}\limits_{\boldsymbol{\theta},\boldsymbol{\vartheta},\boldsymbol{\lambda}}~&~ -\sum_{i=1}^{2}|\mathbf{q}_i^H
\boldsymbol{\theta}|^2+\frac{1}{2\rho}\|\boldsymbol{\theta}-\boldsymbol{\vartheta} +\rho\boldsymbol{\lambda}\|_2^2 \label{OP2}\\
\text {s.t.}~&~\sum_{i=1}^{2}|{\mathbf{h}}_i^H
\boldsymbol{\theta}|^2\leq \gamma, \label{ccc1}\\
~&~|\theta_n|\leq 1, n=1,\cdots,N, \label{ccc3}\\
~&~|{\vartheta}_n|= 1, n=1,\cdots,N, \label{ccc3}
\end{align}
\end{subequations}
where $\rho$ is the penalty parameter and $\boldsymbol{\lambda}$ is the dual variable corresponding to the equality constraint $\boldsymbol{\theta}=\boldsymbol{\vartheta}$. To solve the problem ($\mathcal{P}_7$), we use inner iterations to update $\boldsymbol{\theta}$ and $\boldsymbol{\vartheta}$, and outer iterations to update the dual variables $\boldsymbol{\lambda}$ with the details given below.

For the inner loop, we fix $\boldsymbol{\lambda}$ as a constant. Then, we partition the remaining optimization variables in $(\mathcal{P}_7)$ into two blocks, i.e., $\boldsymbol{\vartheta}$ and $\boldsymbol{\theta}$, and optimize them iteratively as follows. Specifically, when $\boldsymbol{\vartheta}$ is given, $\boldsymbol{\theta}$ can be  updated by solving the following problem.
\begin{subequations}
\begin{align}
(\mathcal{P}_8): \mathop{\min}\limits_{\boldsymbol{\theta}}~&~ -\sum_{i=1}^{2}|\mathbf{q}_i^H
\boldsymbol{\theta}|^2+\frac{1}{2\rho}\|\boldsymbol{\theta}-\boldsymbol{\vartheta} +\rho\boldsymbol{\lambda}\|_2^2 \label{OP3}\\
\text {s.t.}~&~\sum_{i=1}^{2}|{\mathbf{h}}_i^H
\boldsymbol{\theta}|^2\leq \gamma, \label{cd1}\\
~&~|\theta_n|\leq 1, n=1,\cdots,N. \label{ccc3}
\end{align}
\end{subequations}
Note that the objective function of ($\mathcal{P}_8$) is the difference of two non-negative convex functions, which is non-convex in general. To circumvent this difficulty, we approximate the convex function
$|\mathbf{q}_i^H\boldsymbol{\theta}|^2$ by its first-order Taylor expansion at a given point ${\tilde{\boldsymbol{\theta}}}$, denoted by $\underline{q}_i(\boldsymbol{\theta},{\tilde{\boldsymbol{\theta}}})$, i.e.,
\begin{align}\label{taylor1}
|\mathbf{q}_i^H\boldsymbol{\theta}|^2\geq\underline{q}_i(\boldsymbol{\theta}, {\tilde{\boldsymbol{\theta}}}) \triangleq 2 \Re(\bm{\varepsilon}_i^H\boldsymbol{\theta})+\bar{\xi}_i,
\end{align}
where $\bm{\varepsilon}_i\triangleq\mathbf{q}_i^H {\tilde{\boldsymbol{\theta}}}\mathbf{q}_i^H$ and $\bar{\xi}_i=-|\mathbf{q}_i^H{\tilde{\boldsymbol{\theta}}}|^2$.
Therefore, we obtain the following problem,
\begin{subequations}
\begin{align}
(\mathcal{P}_9): \mathop{\min}\limits_{\boldsymbol{\theta}}&~ -2 \sum_{i=1}^{2}\Re(\bm{\varepsilon}_i^H\boldsymbol{\theta})-\sum_{i=1}^{2}\bar{\xi}_i+\frac{1}{2\rho}\|\boldsymbol{\theta}-\boldsymbol{\vartheta} +\rho\boldsymbol{\lambda}\|_2^2 \label{OlP3}\\
\text {s.t.}&~\sum_{i=1}^{2}|{\mathbf{h}}_i^H
\boldsymbol{\theta}|^2\leq \gamma, \label{cld1}\\
&~|\theta_n|\leq 1, n=1,\cdots,N. \label{cl3}
\end{align}
\end{subequations}
Problem $(\mathcal{P}_9)$ is a convex optimization problem, which can
be solved efficiently via existing software, e.g., CVX \cite{cvx}.
By repeatedly solving $(\mathcal{P}_9)$, and setting the point for Taylor expansion,
i.e., ${\tilde{\boldsymbol{\theta}}}$, in each iteration as the optimal solution obtained in the previous iteration, the above procedure generates a sequence of solutions that converge to a Karush-Kuhn-Tucker (KKT) solution of $(\mathcal{P}_8)$ \cite{KKT}.

On the other hand, when $\boldsymbol{\theta}$ is given, $\boldsymbol{\vartheta}$ can be updated by solving the following problem,
\begin{subequations}
\begin{align}
(\mathcal{P}_{10}): \mathop{\min}\limits_{\boldsymbol{\vartheta}}&~ \|\boldsymbol{\theta}-\boldsymbol{\vartheta} +\rho\boldsymbol{\lambda}\|_2^2 \label{OP8}\\
\text {s.t.}&~|{\vartheta}_n|= 1, n=1,\cdots,N, \label{cdd3}
\end{align}
\end{subequations}
for which the optimum is attained when the elements of $\boldsymbol{\vartheta}$ satisfy the following condition,
\begin{align}\label{vn}
{\vartheta}_n^*=\angle({\theta}_n+\rho{\lambda}_n).
\end{align}
In summary, the inner loop alternatively updates $\boldsymbol{\vartheta}$ and $\boldsymbol{\theta}$ until convergence is reached.
\begin{figure}[t]
\centering
\setlength{\abovecaptionskip}{0.cm}
\includegraphics[width=3.5in]{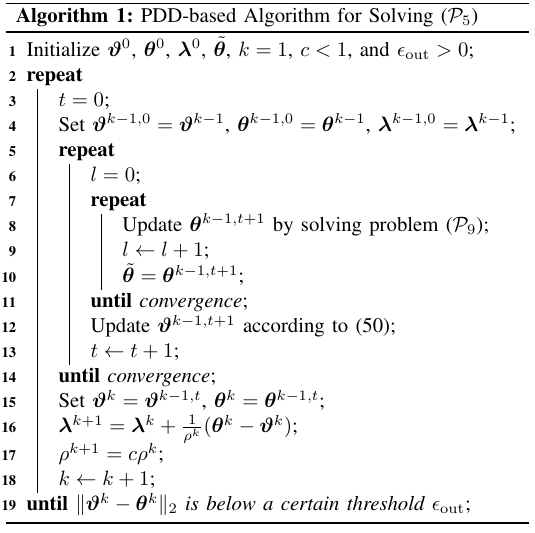}
\label{alg:admm}
\end{figure}

Next, we consider the outer loop, where the dual variable $\boldsymbol{\lambda}$ can be updated by
\begin{align}
\boldsymbol{\lambda}:=\boldsymbol{\lambda}+\frac{1}{\rho}(\boldsymbol{\theta}-\boldsymbol{\vartheta}). \label{OP8}
\end{align}
The above PDD-based algorithm  for solving ($\mathcal{P}_5$) is summarized in Algorithm 1. In step 17 of Algorithm 1, $c < 1$ is a positive constant scaling factor that is used to enlarge the value of the penalty term involved in problem ($\mathcal{P}_7$) in each outer iteration.

Finally, we compare the complexity of the proposed PDD-based method and SDR-based method in \cite{sdr}. By carefully checking the iterative process in Algorithm 1, its computational complexity is found to be dominated by that for solving problem ($\mathcal{P}_9$), which has a complexity of order $\mathcal{O}(N^2)$.
Therefore, the overall complexity of Algorithm 1 can be shown in the order of  $\mathcal{O}(J_0J_1J_2N^2)$, where $J_0$ and $J_1$ denote the maximum outer and inner iteration numbers, and $J_2$ is the number of iterations for solving ($\mathcal{P}_9$). In contrast, the computational complexity of the SDR-based method can be shown in the order of $\mathcal{O}(N^{6.5})$ \cite{sdr}. Thus, the complexity of the proposed Algorithm 1 is much lower than the SDR-based method, especially for large-size IRS with large values of $N$.

\section{Numerical Results}
In this section, we evaluate the performance of our proposed
target-mounted IRS enabled secure wireless sensing design
via numerical results. In the following simulations, the radar pulse waveforms of LRS and URS are given by $x(t)=\exp(j\pi B t^2/t_L)$ and $s(t)=\exp(j\pi B t^2/t_U)$, respectively, with the signal band-width $B=100$ MHz \cite{zeng_wave}. For simplicity, we consider uniform linear array (ULA) for all the LRS, URS, and IRS, where the numbers of antennas at the LRS and URS are $M=64$ and $D=64$, respectively, the numbers of IRS reflecting elements and sensors are $N= 64$ and $L=15$, respectively, the wavelength is $\lambda=0.2$ m, the target is located at $0^\circ$ horizontally from the IRS, and the transmit power of LRS and URS are set as $P_{L} =P_{U} = 30$ mW, if not specified otherwise.
Additionally, the LRS-target and the URS-target distances are set as $d_{\mathrm{LT}}=30$ m and $d_{\mathrm{UT}}=20$ m, respectively; the LRS/URS PRI is $T=100$ us with the pulse durations of $t_L=25$ us and $t_U=30$ us; the total physical size of the IRS is $S=N{d}^{2}$; the target echo surface is set as $A=1.2\times S$ (i.e., the size of target echo surface is larger than that of the IRS); the RCS of the target is set as $\kappa_{\mathrm{t}}=\frac{4\pi A^2}{\lambda^2}$ in m$^2$; and the IRS element separation is ${d}=\lambda/10$, unless otherwise stated. For the purpose of exposition, we first assume that the LRS/URS channel and waveform parameters are perfectly estimated by the IRS sensors, while the effects of their  imperfect estimation on the performance are then evaluated.

For comparison, we consider the following benchmark
schemes:
\begin{itemize}
    \item [1)] {\bf  Random Phase Shift}: We assume that each IRS reflecting
element takes a random phase shift independently
and uniformly distributed in $[0,2\pi)$, based on which we
obtain the received signal power at LRS or URS.

    \item [2)] {\bf No Target-Mounted IRS}: There is no IRS mounted
on the target, as in the traditional sensing system.

    \item [3)] {\bf SDR-based Method}: Problem ($\mathcal{P}_5$) is first  transformed into an semidefinite programming (SDP) problem by applying SDR and dropping the nonconvex rank-one constraint, i.e., $\mathrm{rank}(\boldsymbol{\theta}\boldsymbol{\theta}^H)=1$.
The resulting convex SDP is then optimally solved by CVX \cite{sdr} with 10000 Gaussian randomizations generated to find the best rank-1 feasible solution for $(\mathcal{P}_5)$.
\end{itemize}

\subsection{LRS Present Only }
First, we consider the special case where only the LRS is present in the system to detect the target. In this case, there is no security concern and our objective is to maximize the power of reflected echo signal from the target/IRS towards the LRS receiver for enhancing its target sensing performance, by optimizing the IRS reflection. This problem can be formulated  by setting $\mathbf{q}_1=\sqrt{{{Q_{LS}^2}}}\mathbf{u}$, $\mathbf{q}_2=\mathbf{h}_1=\mathbf{h}_2=\mathbf{0}$ in problem ($\mathcal{P}_5$).
It can be easily shown that the optimal solution to this problem is given by $\boldsymbol{\theta}_{L}^o=\mathbf{u}$. This implies that the optimal IRS reflection should align the reflected signal from IRS with its AoA to IRS for maximizing the signal power at the LRS receiver to
improve its target detection/estimation accuracy.
\begin{figure}[t]
\centering
\setlength{\abovecaptionskip}{0.cm}
\includegraphics[width=3.5in]{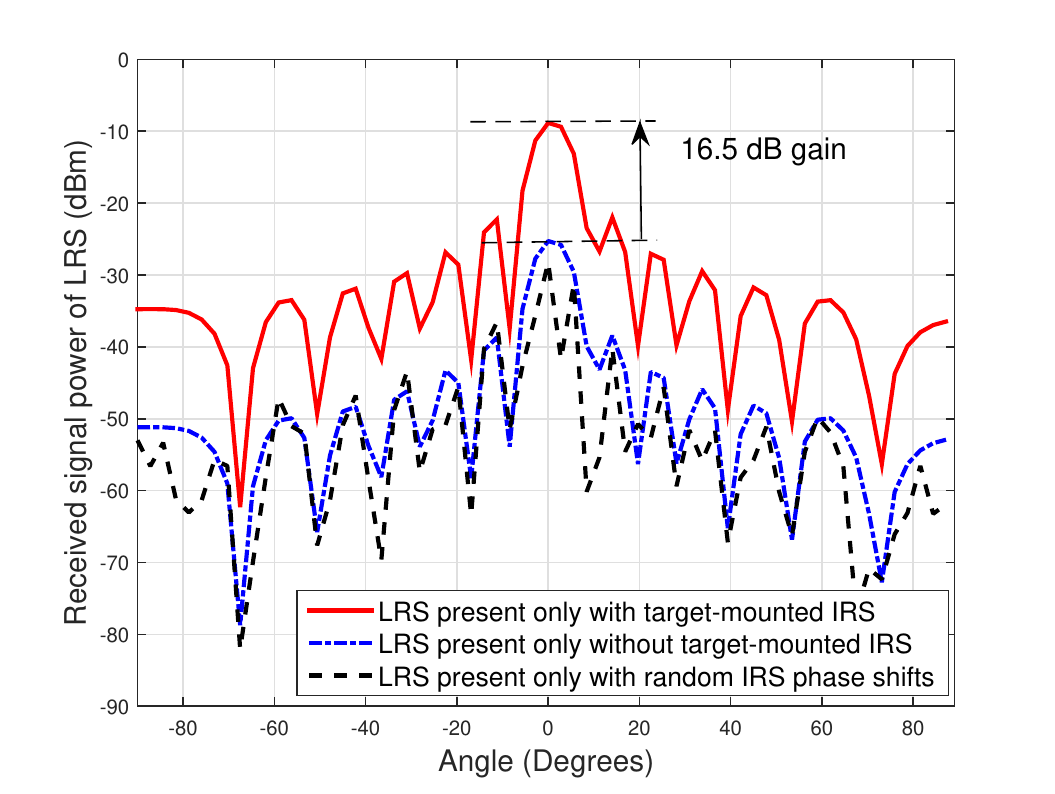}
\caption{Received signal power at LRS versus LRS transmit beam direction.}
\label{NSR_BEAM}
\end{figure}

\begin{figure}[t]
\centering
\setlength{\abovecaptionskip}{0.cm}
\includegraphics[width=3.5in]{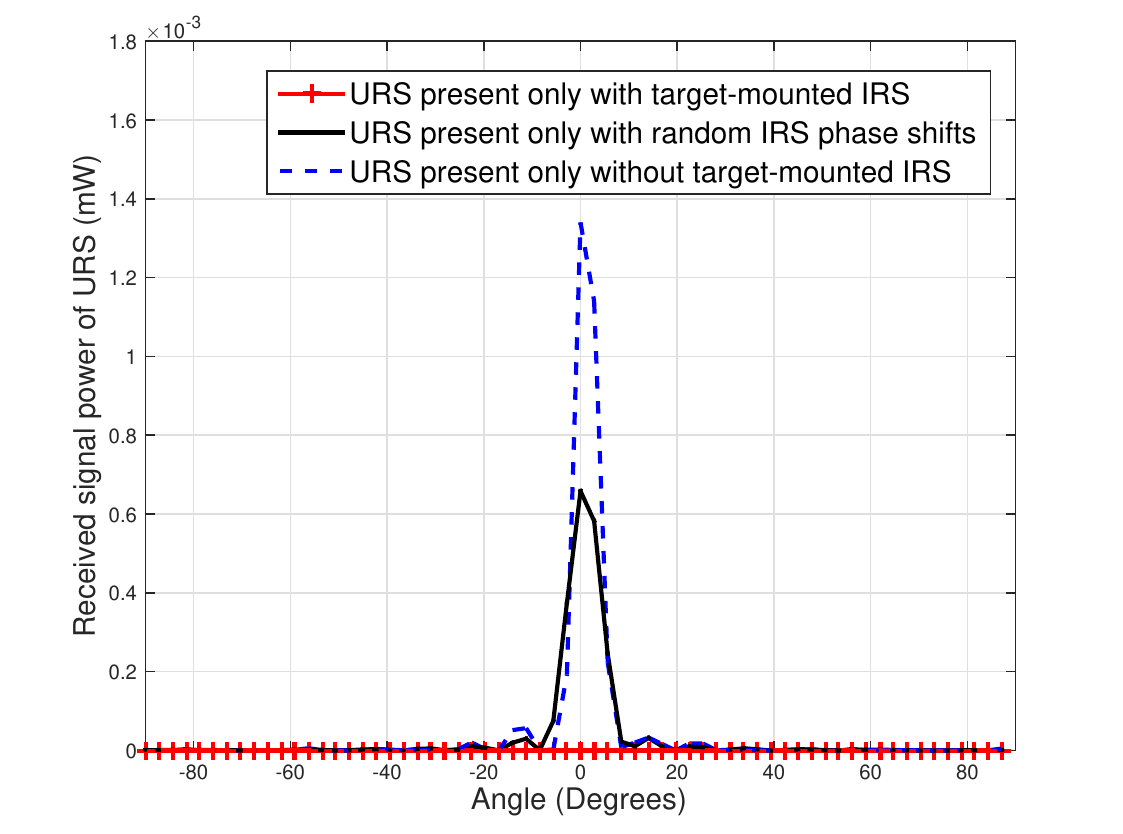}
\caption{Received signal power at URS versus URS transmit beam direction.}
\label{URS_only}
\end{figure}

Fig. \ref{NSR_BEAM} compares the received signal power at the LRS receiver under different schemes versus the LRS transmit beam direction (i.e., by setting different $\mathbf{w}_L$). In this
simulation, we keep the location of the target
fixed, while the LRS transmitter scans different angles in the target
space via transmit beamforming based on the discrete
Fourier transform (DFT) of size $N$. First, it is shown that, despite that the IRS has a smaller size than the target's echo surface, the
received signal power at LRS with
target-mounted IRS yields about 16.5 dB gain over
the traditional sensing without IRS at the target
angle. This is because when the LRS beam
hits the direction of the target, the target-mounted IRS
provides a strong passive beamforming gain for
the echo signal, which is not available for the target
without IRS. Second, with target-mounted IRS, the proposed reflection design  achieves significantly improved signal power at the LRS receiver over the scheme with random IRS phase shifts, which is lack of IRS
passive beamforming gain. Third, it is observed that the scheme without target-mounted IRS performs better than the scheme with random IRS phase shifts. This is because, although the reflected signals in both schemes are not directional to the LRS, the size of the target echo surface is larger than that of the IRS, thus resulting in more reflected signal power at the LRS receiver on average.

\subsection{URS Present Only }
Next, we consider another special case where only the URS is present in the system to detect the same target. In this case, we aim to minimize the power of echo signal from the target/IRS towards the URS receiver by optimizing the IRS reflection. Accordingly, the optimization problem can be formulated as
\begin{align}\label{uro}
\mathop{\min}\limits_{\boldsymbol{\theta}}~&~
\underbrace{|{\mathbf{g}}^H
\boldsymbol{\theta}|^2}_{\text{URS}\rightarrow \text{IRS}\rightarrow \text{URS}}
\nonumber\\
\text {s.t.}~&~|{\theta}_n|= 1, n=1,\cdots,N.
\end{align}
By exploiting the structure of LoS-based channel vector $\mathbf{g}$ given in \eqref{go}, we show in the following proposition a closed-form optimal solution to the above problem.
\begin{figure}[t!]
	\centering
	\subfigbottomskip=2pt
	\subfigcapskip=-5pt
	\subfigure[$N=64$]{
		\includegraphics[width=0.95\linewidth]{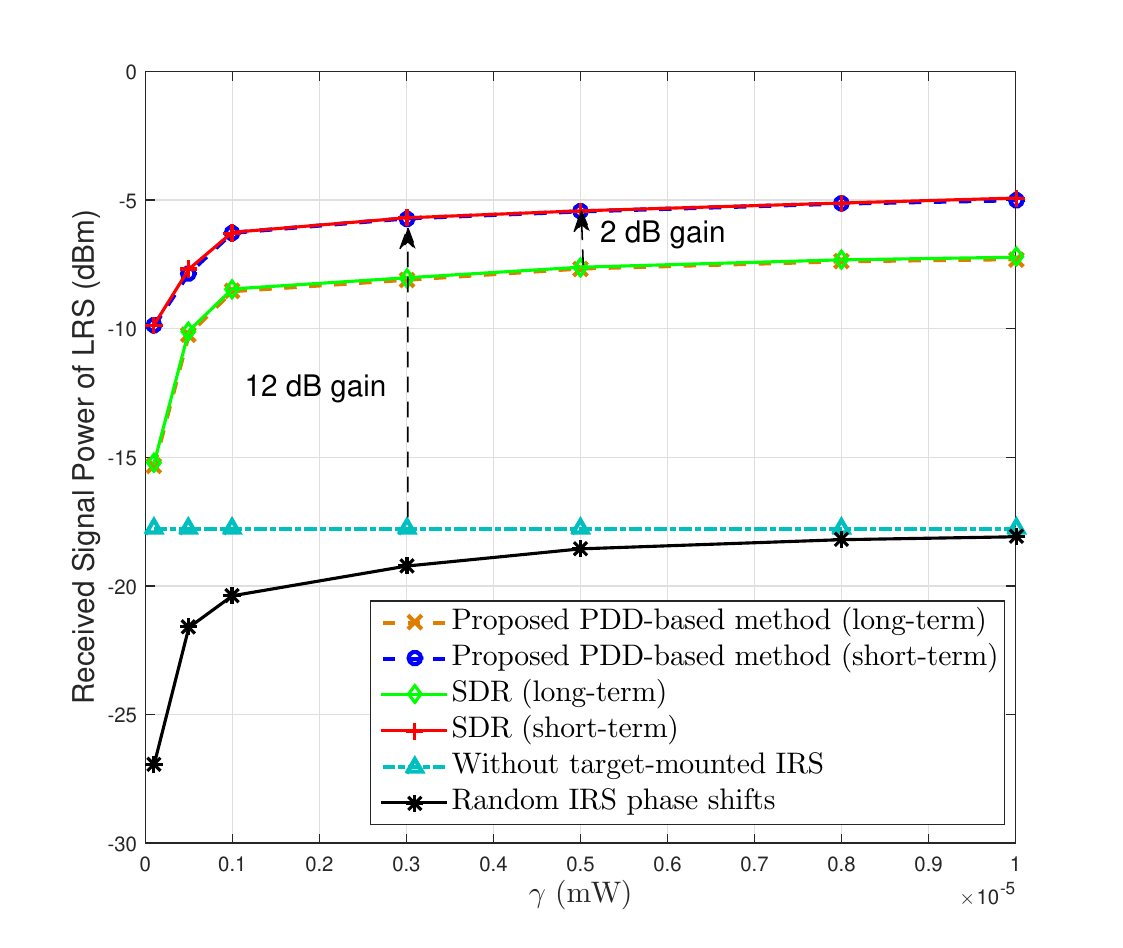}}
	\subfigure[$N=128$]{
		\includegraphics[width=0.95\linewidth]{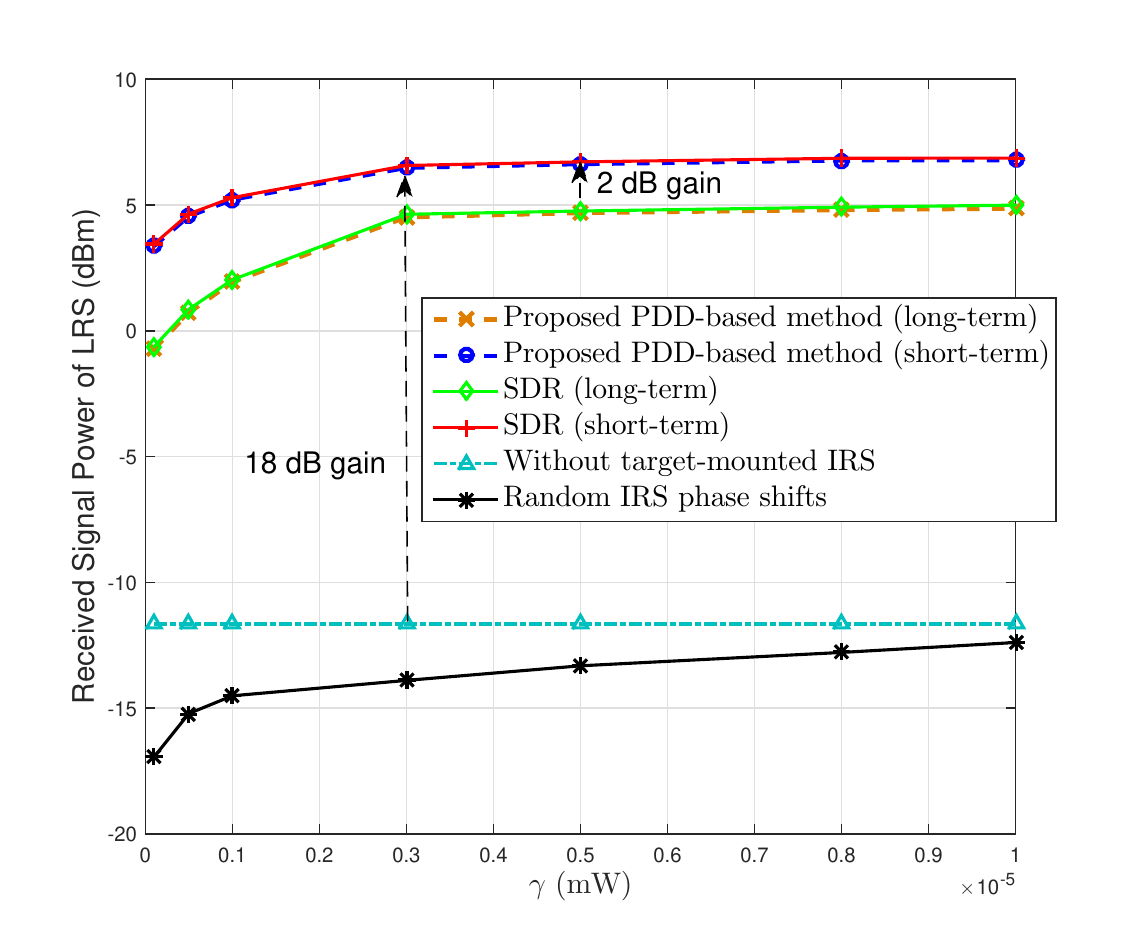}}
	\caption{Received signal power at LRS versus received signal power threshold $\gamma$ at URS.}
\label{gain_N}
\end{figure}

{\bf{Proposition 1}}: The optimal solution to problem \eqref{uro} is given by
\begin{align}\label{so}
\boldsymbol{\theta}_U^o=\boldsymbol{\theta}_x^o\otimes\boldsymbol{\theta}_y^o,
\end{align}
with $\boldsymbol{\theta}_x^o=\mathbf{d}(N_x,\varsigma_x^o)$ and $\boldsymbol{\theta}_y^o
=\mathbf{d}(N_y,\varsigma_y^o)$, where
$\varsigma_x^o=(\zeta_u^{\bar{a}}-\zeta_u^{a})+\frac{\lambda}{N_x{d}}i_x,~i_x=1,2,\cdots,N_x-1$ and $\varsigma_y^o=(\zeta_u^{\bar{e}}-\zeta_u^{e})+\frac{\lambda}{N_y{d}}i_y,~i_y=1,2,\cdots,N_y-1$.
\begin{proof}
Please refer to Appendix A.
\end{proof}

Fig. \ref{URS_only} compares the received signal power at URS receiver under different schemes versus the URS transmit beam direction (similarly assumed as that for the LRS transmitter in Fig. \ref{NSR_BEAM}). We observe that, at the target angle, the received signal power at URS with
target-mounted IRS has a significant reduction compared to both traditional sensing without IRS and the scheme with random IRS phase shifts, which thus validates the effectiveness of target-mounted IRS for achieving secure wireless sensing against the URS. This is because the
target-mounted IRS can completely null the reflected URS signal towards its receiver with the proposed IRS reflection design given in Proposition 1.

\subsection{Both LRS and URS Present}
\begin{figure}[t]
\centering
\setlength{\abovecaptionskip}{0.cm}
\includegraphics[width=3.5in]{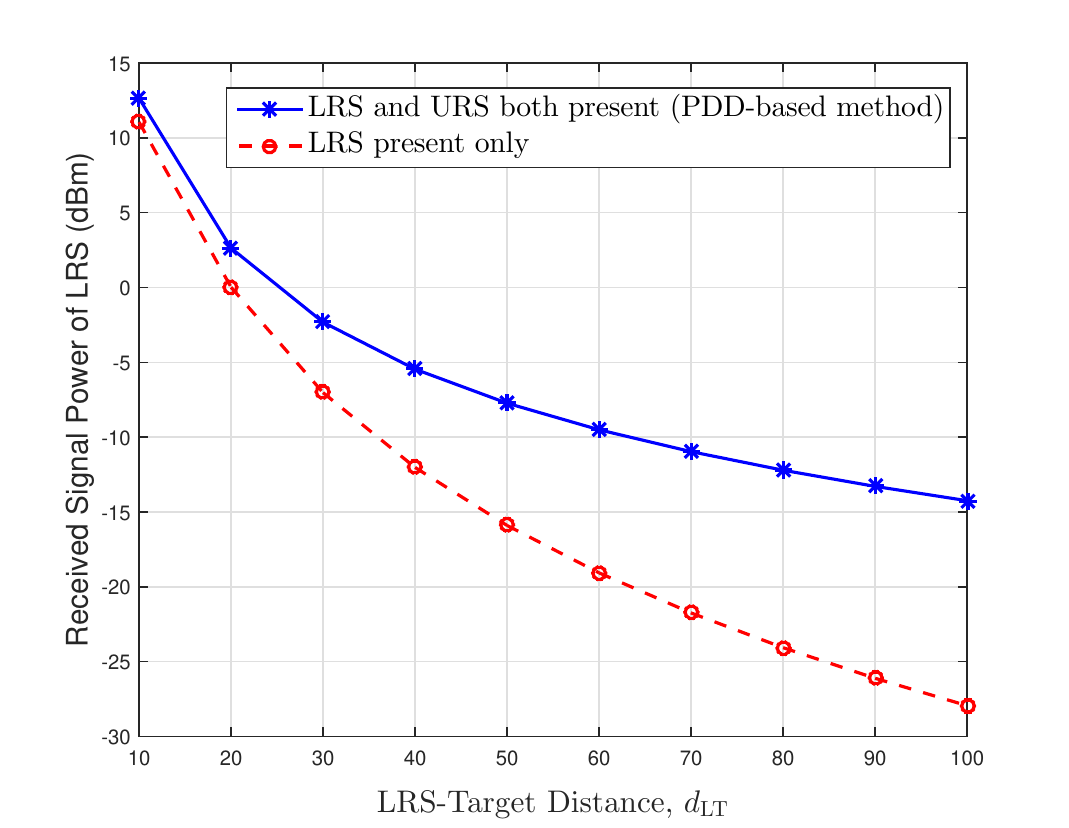}
\caption{Comparison of LRS received signal power with LRS and URS both present versus with LRS present only.}
\label{pro_LRSO}
\end{figure}

Last, we consider the general case where both the LRS and URS are present in the system. In this case, we assume that both the LRS and URS transmitters scan the target space based on DFT codebook, and consider the most challenging scenario when their beam directions both align with the target direction at the same time (otherwise, the system reduces to the LRS-present only or URS-present only as considered previously). Fig. \ref{gain_N} (a) plots the received signal power of LRS versus the maximum received signal power threshold $\gamma$ at the URS. It is observed that the received signal power at LRS is higher when using short-term IRS reflection compared to long-term IRS reflection, which is consistent with the result in \eqref{eq}. In addition, Fig. \ref{gain_N} (a) shows that our proposed PDD-based algorithm can achieve almost the same performance as the SDR-based algorithm, thus being more practically appealing given its much lower computational complexity. Moreover, the proposed short-term and long-term IRS reflections both achieve significantly improved received signal power at LRS over the benchmark schemes without target-mounted IRS or with random IRS phase shifts. In Fig.\ref{gain_N} (b), we increase the number of IRS reflecting elements from $N=64$ to $N=128$ (by decreasing the IRS element separation ${d}$) with the physical size of the IRS (i.e., $S$) fixed. It is observed that the performance gap between our proposed short-term/long-term IRS reflection and the scheme without target-mounted IRS is enlarged as $N$ increases. This is because installing more reflecting elements at IRS provides enhanced passive beamforming gain.

Next, we present a performance comparison of the proposed scheme with both LRS and URS present ($\gamma=10^{-5}$ mW) versus the LRS-present-only scenario. For both cases, the IRS reflection is optimized as above. As
shown in Fig. \ref{pro_LRSO}, the URS-target distance is kept constant, i.e., $d_{UT}=20$ m while the LRS gradually moves away from the target. It is observed that the proposed scheme shows better performance in terms of LRS received signal power when both LRS and URS are present, as compared to the scenario where only LRS is present, and their performance gap is more pronounced for a larger LRS-target distance $d_{LT}$.
This demonstrates the advantage of using URS radar signals for the LRS target's detection, especially when the LRS is located far away from the target and its own radar signal is severely attenuated due to the increased LRS-target distance.
\begin{figure}[t]
\centering
\setlength{\abovecaptionskip}{0.cm}
\includegraphics[width=3.5in]{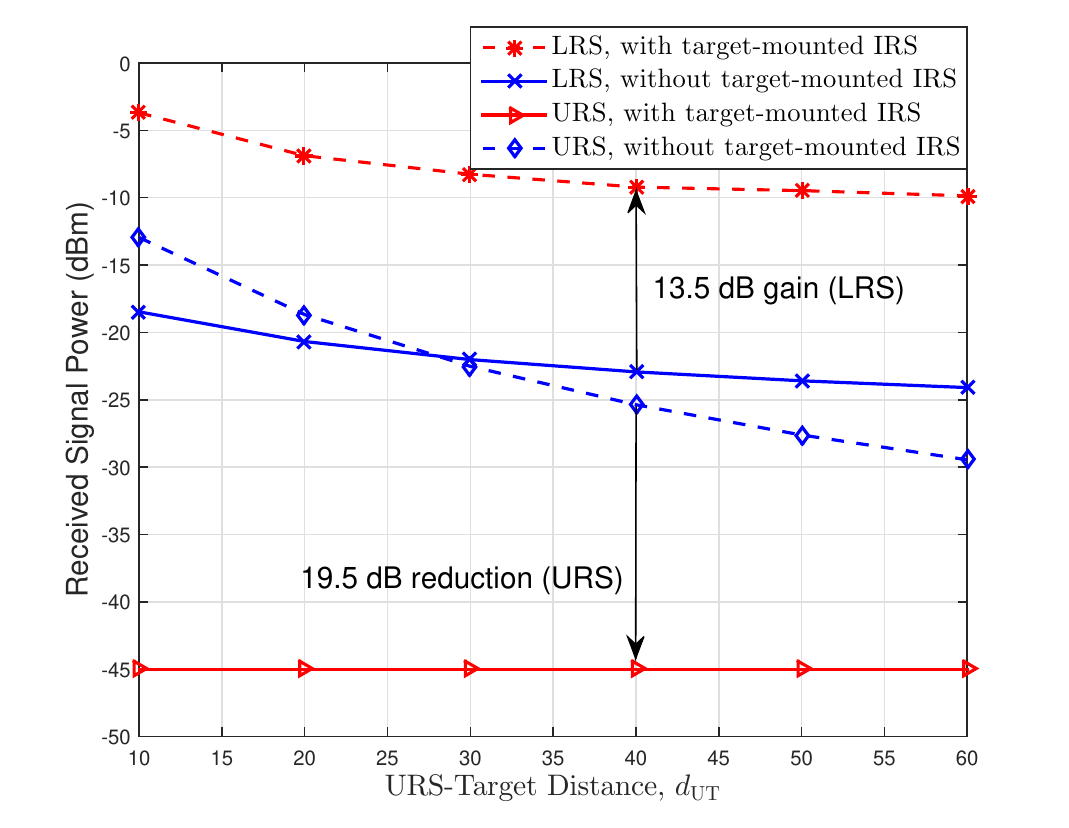}
\caption{The performance of the proposed scheme versus  URS-target distance.}
\label{distance_short}
\end{figure}

\begin{figure}[t]
\centering
\setlength{\abovecaptionskip}{0.cm}
\includegraphics[width=3.5in]{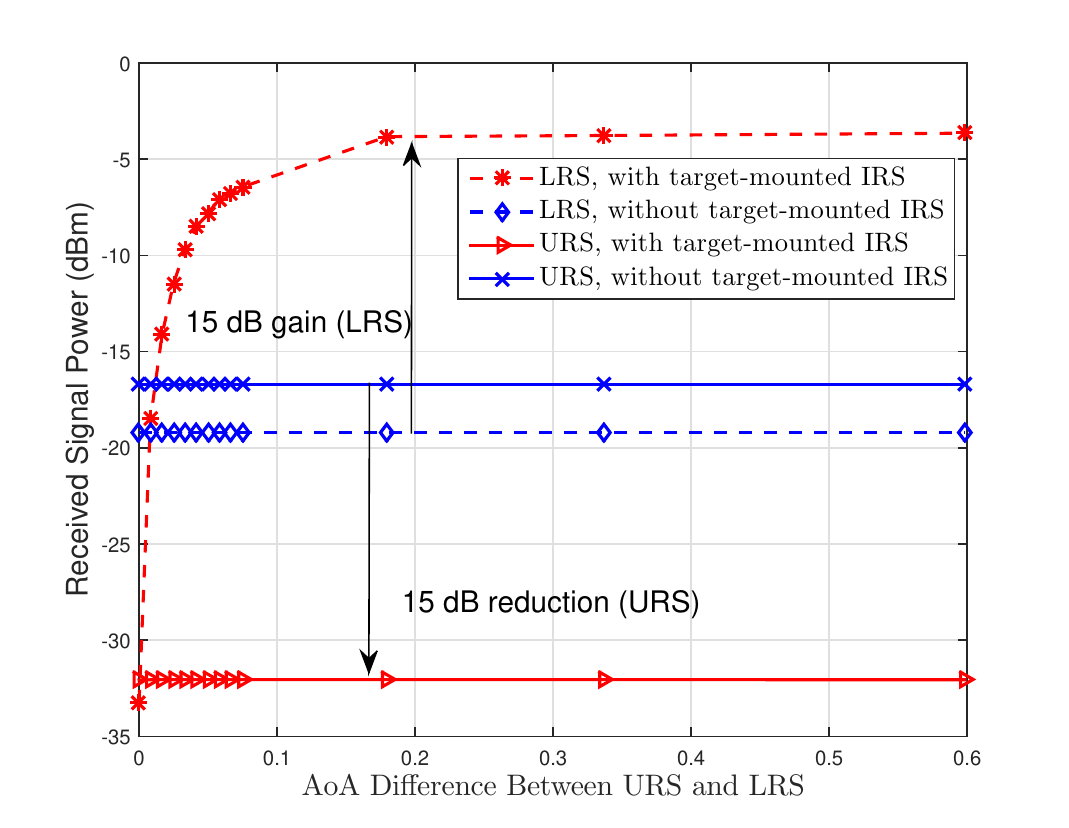}
\caption{The performance of the proposed scheme versus the AoA difference between URS and LRS.}
\label{angle_short}
\end{figure}

In Fig. \ref{distance_short}, we keep the AoAs from LRS and URS to target/IRS fixed, and plot the received signal powers at LRS and URS with or without target-mounted IRS versus the distance between the URS and target. For the case with target-mounted IRS, the proposed PDD-based algorithm is applied with short-term IRS reflection. It is observed that compared to the case without target-mounted IRS, the proposed design with target-mounted IRS can significantly enhance the received signal power at LRS and reduce that at URS at the same time, for a large range of URS-target distance values.

In Fig. \ref{angle_short}, we keep LRS-target distance and URS-target distance fixed, and plot the received signal powers at LRS and URS with or without target-mounted IRS versus the AoA difference between LRS and URS. It is shown that for the scheme without target-mounted IRS, the received powers at LRS and URS are constant due to fixed LRS-target and URS-target distances, regardless of their AoAs to the target/IRS. In contrast, the proposed design with target-mounted IRS can achieve greatly enhanced/suppressed power at LRS/URS receivers, provided that the AoA difference between LRS and URS is sufficiently large, say, larger than 0.1 radian (rad).
\begin{figure}[t]
\centering
\setlength{\abovecaptionskip}{0.cm}
\includegraphics[width=3.5in]{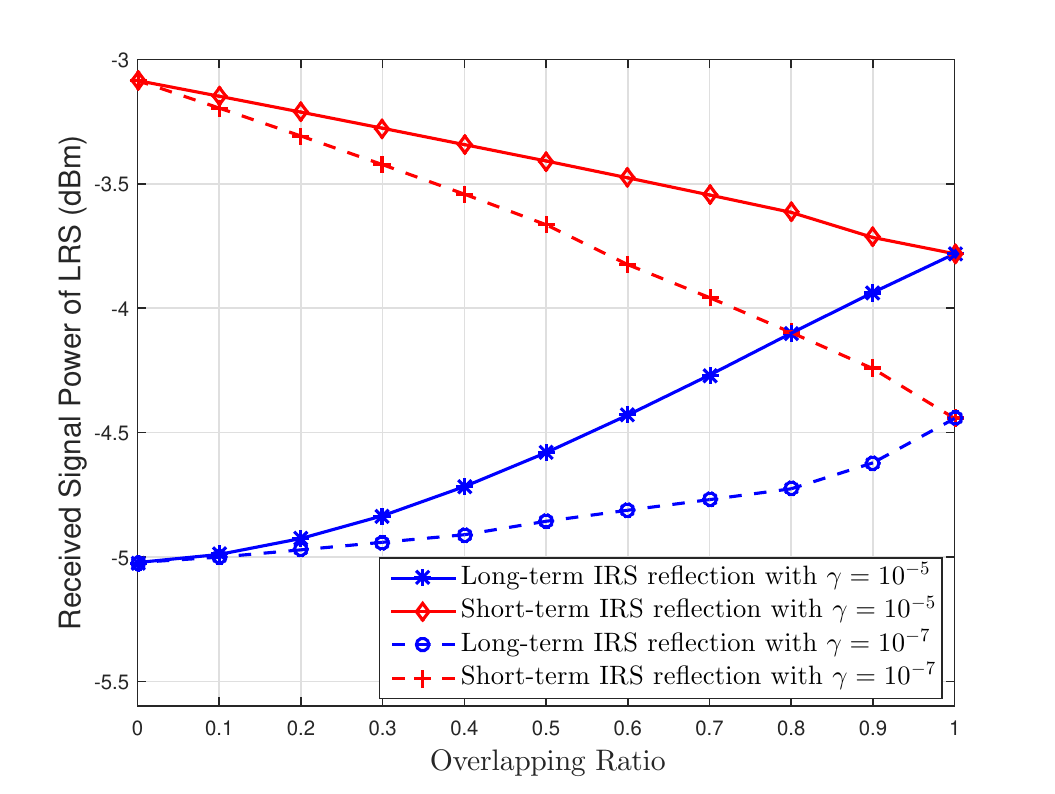}
\caption{Received signal power at LRS versus the LRS/URS signal overlapping  ratio with short-term or long-term IRS reflection.}
\label{overlapped_ratio}
\end{figure}

\begin{figure}[t]
\centering
\setlength{\abovecaptionskip}{0.cm}
\includegraphics[width=3.5in]{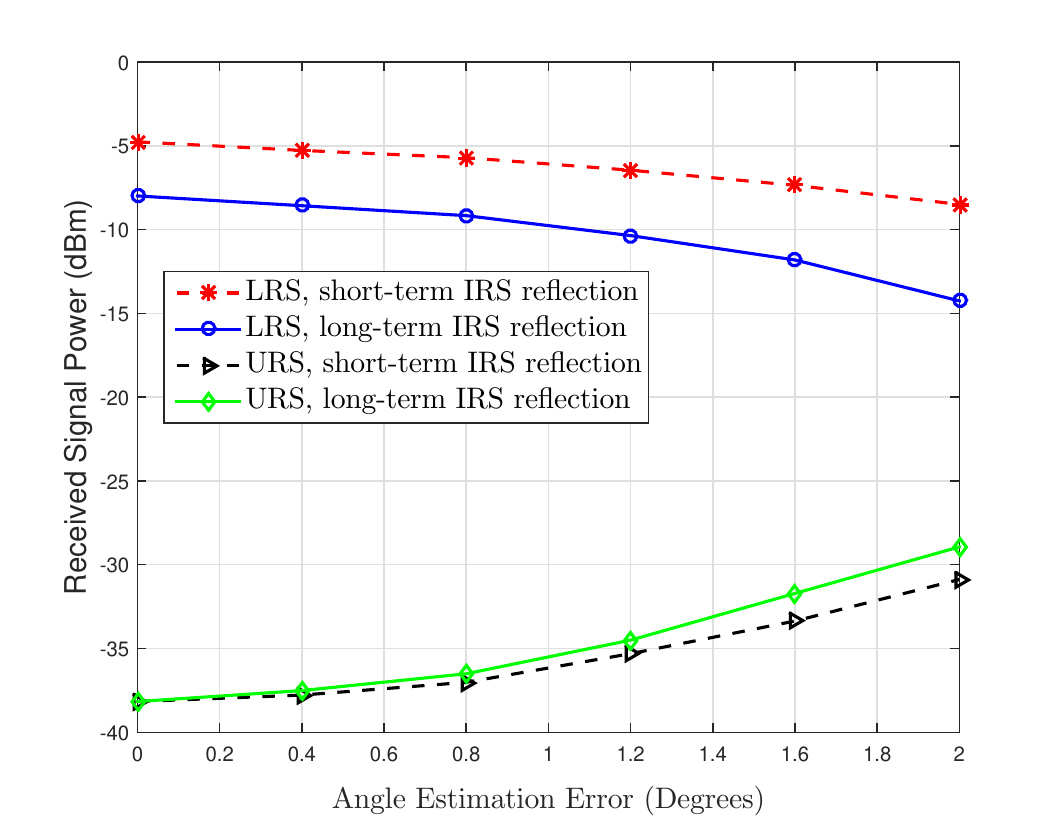}
\caption{Effect of imperfect IRS sensor angle estimation to LRS/URS received signal power with short-term or long-term IRS reflection.}
\label{error}
\end{figure}

Fig. \ref{overlapped_ratio} compares the performance of long-term versus short-term IRS reflection for the proposed target-mounted IRS design under different ratios of the LRS/URS overlapping signal duration to that of the LRS/URS pulse duration, i.e., $t_o/t_U$ by assuming $t_L=t_U=30$ us. It is observed that the received signal power at LRS with long-term IRS reflection is upper-bounded by that with short-term IRS reflection, as expected. In addition, as the LRS/URS signal overlapping ratio increases, the received signal power at LRS with long-term IRS reflection increases, while that with short-term IRS reflection decreases; and as the overlapping ratio approaches one, they become identical, which is consistent with the discussion for  \eqref{eq}. Furthermore, it is observed that more stringent security requirement (i.e., smaller $\gamma$) on the URS received signal power also results in lower received signal power at LRS. This is because the IRS reflection needs to further reduce the signal power towards the URS, which has to sacrifice the passive beamforming gain towards the LRS.

Finally, we evaluate the impact of imperfect LRS and URS
AoA estimation by IRS sensors on the received signal powers at LRS and URS
for the proposed target-mounted IRS design, as shown in Fig. \ref{error}. It is observed that the proposed scheme with
imperfect AoA estimation has less received signal power at
LRS and more received signal power at URS, as compared to
the case with perfect AoA estimation. Although inaccurate
AoA estimation reduces the signal enhancement/suppression
gain of target-mounted IRS in practice, the performance loss
is observed to be quite small (less than 1.8\%) if the angle estimation error is below $1^{\circ}$.

\section{Conclusions}
In this paper, we proposed a new secure wireless sensing approach to simultaneously enhance the target detection for LRS while preventing that against the URS by utilizing the target-mounted IRS.
We presented a practical protocol for target-mounted IRS and designed the IRS reflection to maximize the IRS reflected
signal power to LRS, while keeping that at URS
below a certain threshold, for both short-term and long-term IRS reflection scenarios. Simulation results showed the significant received signal power improvement/reduction at LRS/URS achieved with our proposed
target-mounted IRS designs as compared to
benchmark schemes, and also validated their robust performance
under different practical setups.

\begin{appendices}
\section{The Proof of \eqref{so}}
The IRS reflection gain $|{\mathbf{g}}^H
\boldsymbol{\theta}|^2$ in \eqref{uro} can be decoupled as the
product of the gains along the $x$- and $y$-axes, i.e.,
\begin{align}\label{hv}
|{\mathbf{g}}^H\boldsymbol{\theta}|^2=|\mathbf{d}^H(N_x,\zeta_u^{\bar{a}}-\zeta_u^{a})
\boldsymbol{\theta}_x|^2\times|\mathbf{d}^H(N_x,\zeta_u^{\bar{e}}-\zeta_u^{e})
\boldsymbol{\theta}_y|^2.
\end{align}
By setting the IRS reflection vector $\boldsymbol{\theta}$ along the $x$- and $y$-axes as $\boldsymbol{\theta}_x=\mathbf{d}(N_x,\varsigma_x)$ and $\boldsymbol{\theta}_y=\mathbf{d}(N_y,\varsigma_y)$, respectively,  with $\varsigma_x$ and $\varsigma_y$ being the corresponding steering angles, we have
\begin{align}
|\mathbf{d}^H(N_x,\zeta_u^{\bar{a}}-\zeta_u^{a})
\boldsymbol{\theta}_x|^2&=\left|\sum_{n=1}^{N_x}e^{j\frac{2\pi d}{\lambda}(n-1)(\varsigma_x-(\zeta_u^{\bar{a}}-\zeta_u^{a}))}\right|^2\nonumber\\
&=\frac{\sin^2(\frac{\frac{2\pi d}{\lambda}N_x\delta_x}{2})}{\sin^2(\frac{\frac{2\pi d}{\lambda}\delta_x}{2})},\label{hi}\\
|\mathbf{d}^H(N_y,\zeta_u^{\bar{e}}-\zeta_u^{e})
\boldsymbol{\theta}_y|^2&=\left|\sum_{n=1}^{N_y}e^{j\frac{2\pi d}{\lambda}(n-1)(\varsigma_y-(\zeta_u^{\bar{e}}-\zeta_u^{e}))}\right|^2\nonumber\\
&=\frac{\sin^2(\frac{\frac{2\pi d}{\lambda}N_y\delta_y}{2})}{\sin^2(\frac{\frac{2\pi d}{\lambda}\delta_y}{2})},\label{hi1}
\end{align}
where $\delta_x=\varsigma_x-(\zeta_u^{\bar{a}}-\zeta_u^{a})$ and $\delta_y=\varsigma_y-(\zeta_u^{\bar{e}}-\zeta_u^{e})$. Based on \eqref{hi} and \eqref{hi1}, when  $\delta_x=\frac{\lambda}{N_xd}i_x,~i_x=1,2,\cdots,N_x-1$, it follows that $
|\mathbf{d}^H(N_x,\zeta_u^{\bar{a}}-\zeta_u^{a})
\boldsymbol{\theta}_x|^2=0$.
Similarly, when $\delta_y=\frac{\lambda}{N_yd}i_y,~i_y=1,2,\cdots,N_y-1$, we have
$|\mathbf{d}^H(N_y,\zeta_u^{\bar{e}}-\zeta_u^{e})
\boldsymbol{\theta}_y|^2=0$.
Thus, the optimal solution to problem \eqref{uro} is
given by \eqref{so}.
\end{appendices}

\end{document}